\documentclass[a4paper,11pt,leqno]{amsart}
\usepackage{graphicx}
\usepackage{amsmath,amssymb}
\usepackage{floatrow}
\usepackage[longnamesfirst,comma]{natbib}
\numberwithin{equation}{section}
\newtheorem{theorem}{Theorem}

\newtheorem{lemma}[theorem]{Lemma}
\newtheorem{proposition}[theorem]{Proposition}

\newtheorem{definition}[theorem]{Definition}
\numberwithin{theorem}{section}
\numberwithin{figure}{section}
\numberwithin{table}{section}

\newcommand\npar{\medskip\noindent}
\renewcommand\ni{\noindent}

\newcommand\subhead{\noindent\textit}

\newcommand\bE{{\mathbb E}}

\newcommand\bP{{\mathbb P}}

\newcommand\bR{{\mathbb R}}

\renewcommand\d{\partial}

\newcommand\cF{\mathcal{F}}
\newcommand\cB{\mathcal{B}}

\newcommand\cG{\mathcal{G}}

\newcommand\cI{\mathcal{I}}
\newcommand\cP{\mathcal{P}}

\newcommand\sig{\sigma}

\renewcommand\d{\partial}
\newcommand\bstar{\begin{eqnarray*}}
\newcommand\estar{\end{eqnarray*}}
\newcommand\be{\begin{equation}}
\newcommand\ee{\end{equation}}
\newcommand\bea{\begin{eqnarray}}
\newcommand\eea{\end{eqnarray}}
\newcommand\1{{\bf 1}}
\renewcommand\ni{\noindent}
\newcommand\clc{,\ldots ,}
\newcommand\ha{\frac{1}{2}}

\newcommand\fs{\mathfrak{s}}
\newcommand\ma{\mathfrak{a}}
\newcommand\mH{\mathfrak{H}}
\newcommand\mP{\mathfrak{P}}

\newcommand\tY{\tilde{Y}}
\newcommand\tH{\tilde{H}}

\newcommand\aS{\langle S\rangle}

\usepackage{color}

\title{Verification of Internal Risk Measure Estimates}
\thanks{The author thanks Paul Embrechts for introducing him to this topic, Phil Dawid for discussions of prequential statistics, and Johanna Ziegel for extremely helpful comments and advice. He is also grateful to two anonymous referees for their careful reading of the paper and many constructive suggestions for improvements. Some of this work was carried out at the Hausdorff Research Institute for Mathematics at the University of Bonn, where the author was a visitor during the Trimester Program \emph{Stochastic Dynamics in Economics and Finance} in August 2013. Discussions there have been very helpful in clarifying the content of this paper.}
\author{Mark H.A. Davis}
\address{Department of Mathematics, Imperial College London}
\email{mark.davis@imperial.ac.uk}
\date{\today{}.}
\begin{document}
\begin{abstract}
This paper concerns sequential computation of risk measures for financial data and asks how, given a risk measurement procedure, we can tell whether the answers it produces are `correct'. We draw the distinction between `external' and `internal' risk measures and concentrate on the latter, where we observe data in real time, make predictions and observe outcomes.  It is argued that evaluation of such procedures is best addressed from the point of view of probability forecasting or Dawid's theory of `prequential statistics' [Dawid, JRSS(A)1984]. We introduce a concept of `calibration' of a risk measure in a dynamic setting, following the precepts of Dawid's weak and strong prequential principles, and examine its application to quantile forecasting (VaR -- value at risk) and to mean estimation (applicable to CVaR -- expected shortfall). The relationship between these ideas and `elicitability' [Gneiting, JASA 2011] is examined. We show in particular that VaR has special properties not shared by any other risk measure. Turning to CVaR we argue that its main deficiency is the unquantifiable tail dependence of estimators. In a final section we show that a simple data-driven feedback algorithm can produce VaR estimates on financial data that easily pass both the consistency test and a further newly-introduced statistical test for independence of a binary sequence.

\medskip\ni \textsc{JEL classification:} C13\,\,C32\,\,C44\,\,C53\,\,G17

\ni\textsc{Key words:} Risk measures, probability forecasting, prequential statistics, quantile and mean forecasting, consistency of estimates.
\end{abstract}
\maketitle
\section{Introduction}\label{sec:intro}
Computing risk measures is a matter of primary importance to the financial services industry, both from the point of view of short-term risk management and for regulatory capital allocation purposes; see \citet{emho14} for a recent survey. For a portfolio of assets, a risk measure is generally interpreted as some functional of the conditional distribution $F$ of the portfolio loss\footnote{In the literature on risk management there is no settled convention whether losses should be designated as positive or negative. In this paper losses are positive, so we are concerned with the \emph{right-hand} tail of the distribution. Distribution functions will be taken to be right-continuous, i.e. $F_Y(y)=\bP[Y\leq y]$.} between times $t$ and $t+h$ given all the market information up to today, time $t$. The prediction horizon $h$ is typically a week or 10 days for market risk management, somewhat longer for credit-related assets or insurance. The most widely used risk measures are the \emph{value at risk} VaR,  and CVaR, variously known as \emph{conditional value at risk}, \emph{expected shortfall} or \emph{expected loss beyond} VaR. Formal definitions are as follows. The definition of CVaR for general distributions involves some subtleties, for which the reader is referred to \citet{rocury02}.

\begin{definition}\label{defvar}
Let $F$ be a right-continuous distribution function on the real line and $\beta\in(0,1)$.

\noindent(i)The $\beta$-\emph{quantile} is the interval $[q_\beta^-,q_\beta^+)$ if $F(q_\beta^+)>F(q_\beta^+-)$ and the interval $[q_\beta^-,q_\beta^+]$ otherwise, where $q_\beta^-=\inf\{y:F(y)\geq\beta\}$ and $q_\beta^+=\inf\{y:F(y)>\beta\}$.

\noindent (ii) The value at risk at level $\beta$ is $\mathrm{VaR}_\beta=q_\beta^-$.  

\noindent(iii) $\mathrm{CVaR}_\beta$ is the mean of the $\beta$-tail distribution, given by
\[ F_\beta(y)=\left\{\begin{array}{ll}0,&y<q_\beta^-\\
\frac{F(y)-\beta}{1-\beta},&  y\geq q_\beta^-.\end{array}\right.\]
Explicitly,
\bea \mathrm{CVaR}_{\beta}&=&\mathrm{VaR}_\beta+\frac{1}{1-\beta}
\int_{(q_\beta^-,\infty)}(y-q_\beta^-)F(dy)\label{cvar1}\\
&=& \frac{1}{1-\beta}\left[\int_{F(q_\beta^-)}^1q_\tau^-d\tau 
+q_\beta^-(F(\beta)-\beta)\right].\label{cvar2}\eea
\end{definition}

Expression \eqref{cvar1} quantifies the gap between VaR and CVaR, while \eqref{cvar2} provides a relationship between CVaR and quantiles. When there is no jump at the $\beta$-quantile, \eqref{cvar2} reduces to the familiar expression 
\[\mathrm{CVaR}_\beta=\frac{1}{1-\beta}\int_\beta^1 \mathrm{VaR}_\tau d\tau.\]
There has been renewed debate about the relative merits of VaR and CVaR and indeed about the risk management process as a whole. A huge literature on risk measures was triggered off by the seminal paper by \citet*{adeh99} in which axioms for a `coherent' risk measure were formulated, but much of this literature is presented in a pure mathematical framework taking no account of where the data is coming from, how the risk measure is to be computed or what the ultimate purpose of the exercise is. Clearly these practical considerations have to be included in any evaluation of the risk management process. For example, \citet*{cds10} concentrate on stability of computation under perturbations of the model and conclude that computation of CVaR suffers from unavoidable instabilities that are not present in the computation of VaR, challenging the at the time conventional wisdom that CVaR is to be preferred because it is coherent while VaR is not. In a similar vein, \citet{kph13} draw the distinction between `external' and `internal' risk management, which we discuss in Section \ref{erm} below. They give a revised set of axioms appropriate to external risk management and show that value at risk does satisfy these axioms. They make the interesting suggestion of replacing CVaR by  CMVaR, the conditional \emph{median} shortfall. Of course, at threshold level $\beta$, CMVaR is just VaR at level $\ha(1+\beta)$, so CMVaR satisfies the Kou-Peng-Heyde axioms and has all the computational advantages of VaR while making some attempt at quantifying tail risk. Further comments will be found in Section \ref{sec:cvar}.

In this paper we take a different tack. Given that we have selected a risk measure and a computational algorithm, how can we tell whether the answer is `correct' when we apply the algorithm to real data? This is by no means a simple question, for one very clear reason. In evaluating, say, the value at risk at level $\beta$, what we are computing at time $(k-1)$ is $(F^m_k)^{-1}(\beta)$, the $\beta$th quantile of the conditional distribution $F^m_k$ of the portfolio return at time $k$ given all information up to $(k-1)$, computed according to a chosen model, labelled $m$. Even if the model $m$ is time-invariant, $F^m_k$ is a \emph{different distribution} for each $k$, because the conditioning event is different. When time $k$ arrives, we observe \emph{one number}, which may or may not exceed the predicted quantile level. How to evaluate the quality of such predictions is the province of \emph{probability forecasting} \citep{daw86, lsg11, gne11}, a branch of statistics that until very recently was largely ignored by researchers in financial risk management, although some of its techniques are routinely used in various ways by practitioners. The approach we take is inspired by P. Dawid's theory of \emph{prequential probability} \citep{daw84}, and in particular to the  exceptionally stimulating paper \citet{dawvov99}. 

The paper is laid out as follows. In Section \ref{sec:fund} we discuss the fundamentals of time series prediction in general terms, highlighting the special features of financial price data. We also introduce the essential distinction between \emph{external} and \emph{internal} risk measures together with some comments on the applicability of the prequential approach in the latter case. Section \ref{sec:el} defines elicitability from statistical decision theory and summarises the information we need. The core of the paper is Section \ref{sec:cp}, in which Definition \ref{def:cons}  formalizes the concept of \emph{calibration} of a risk management statistic in a dynamic setting, and we demonstrate its connection to elicitability via identification functions. In Section \ref{sec:qf} we examine the case of quantile forecasting and VaR estimation in some detail and show that this case has especially favourable features: the quantile statistic is calibratable under essentially no conditions on the underlying data (Theorem \ref{thm:quantest}). In this section we also introduce an auxiliary test for independence, the details of which are given in Appendix \ref{sec:mcm}, and we also discuss an example, given by \citet{he14}, showing that calibration and independence, while necessary conditions for correct prediction, are far from sufficient. Section \ref{sec:mv} covers mean estimation, including CVaR. The main calibration result here is Theorem \ref{thm:meantest}, utilising martingale convergence theorems. We discuss the CVaR problem in more detail in Section \ref{sec:cvar}; first we describe the \citet{rocury00} characterisation of CVaR as the solution to a minimisation problem and its role as a further verification test for VaR predictors; then we highlight the main problem in this area: unquantifiable tail dependence.  Finally, we illustrate these ideas by looking at one week ahead quantile prediction for the FTSE100 stock index, showing that a simple data-driven algorithm can produce a sequence of quantile forecasts that easily survive calibration and independence tests.

\section{Fundamentals of prediction}\label{sec:fund}
\subsection{Background} Financial risk management is essentially about prediction: given whatever information we think relevant, we have to take a view on the likely returns of a portfolio over some holding period, generally quantified by some risk measure. It may be helpful to start out by placing this problem in the general context of time series prediction. In any such problem, the approach taken must depend on the nature of the data, on what it is we are trying to predict and on what the purpose of the prediction is. There is a hierarchy of possibilities. 

\npar(i) The simplest case is coin tossing: the whole probabilistic structure is fixed axiomatically, with no need for statistical modelling. No-one will quarrel with the statement that the distribution of the number of heads in the next $n$ tosses is the Binomial distribution $B(n,0.5)$. 

\npar(ii) There are situations in which the data is produced by a well-understood and stable physical mechanism, for example Geiger counter detection of radioactive emissions. Here it is clear from the physics that the sequence of counts will constitute a Poisson process. We can estimate its rate from past data, and there is no reason to suppose that the rate will be different in the future, or at most will vary in predictable ways, at least over short time scales.

\npar(iii)\, Next, weather forecasting. This is similar to the previous case (ii) in that predictions are mainly derived from mathematical models describing the underlying physics, but of course the latter are extremely complex; see \citet{war10} for an authoritative account. Prediction is bound up to a great extent with studies of the effects on the models of perturbations in model parameters and/or initial conditions, a subject that has become a discipline in its own right under the name of \emph{Uncertainty Quantification} \citep{smi14}. Other approaches include the more statistical topic of pattern matching, in which periods in the past are identified in which the weather patterns match those of today.

It is useful to distinguish two very different problems in climate science (a) short-term weather forecasts (up to a week) and (b) extreme-value problems such as flood barrier design, estimates of the probability of  inundation of coastal areas or incidence and severity of tornados. Short-term forecasts can be monitored by checking whether the predictions are well 'calibrated'. We give an example below, and indeed the main purpose of this paper is to formalize this idea. Techniques to deal with problems in case (b) are completely different because, by definition, the data is sparse. Flood barrier design is an engineering problem in which complex mathematical models and extreme-value theory \citep{ekm97} are combined to give best-possible estimates of the frequency of exceedance of various high-water marks; this is followed by a cost-benefit analysis in which the level of protection is traded off against the cost of providing it. The point is that we are never in the position of case (a) where a sequence of predictions can be checked against subsequent outcomes. Instead, physics and data are combined to assemble evidence in favour of a decision the consequences of which will, we hope, never be tested to destruction.

\npar(iv) Statistics comes into its own in situations where we have no physical model but an adequate supply of data that is `reasonably predictable'. A case in point would be a sequence of insurance claims resulting from car accidents. There is no physical theory, but there is a huge amount of relevant data. Individual accidents are largely independent, and the general prevalence of accidents and the claims arising from them depend on well-understood factors such as the age distribution of drivers, repair costs, growth in traffic and improvements in safety due to engineering developments, speed restrictions, etc. In these circumstances highly credible statistical models can be built, giving an accurate view of the claims likely to be faced by an insurance company.

\subsection{Financial risk management}\label{finrisk} Now we move to the subject of this paper, the computation of risk management parameters for portfolios of financial assets. The problem could hardly be more different from those in cases (i)-(iii) above. The underlying reason for this is that the economy is a meso-scale phenomenon: too big to be modelled in complete detail, but too small, and too interconnected, to be treated by methods of statistical mechanics.

\begin{figure}[ht]
\centering
\begin{minipage}[b]{0.45\linewidth}
\includegraphics[scale=0.4]{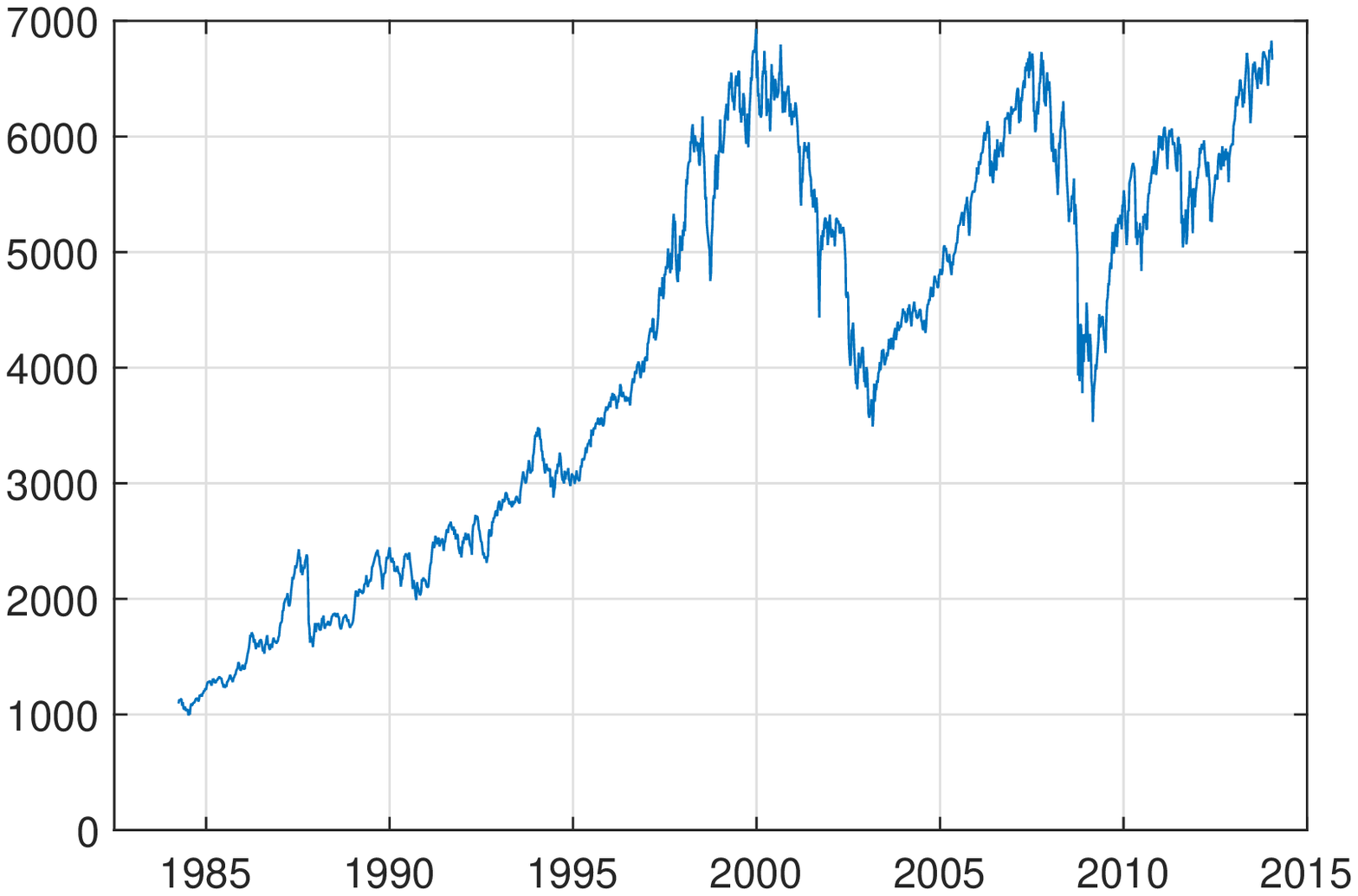}

\vspace{-5mm}
\begin{center}(a) Index values.\end{center}
\end{minipage}
\qquad
\begin{minipage}[b]{0.45\linewidth}
\includegraphics[scale=0.4]{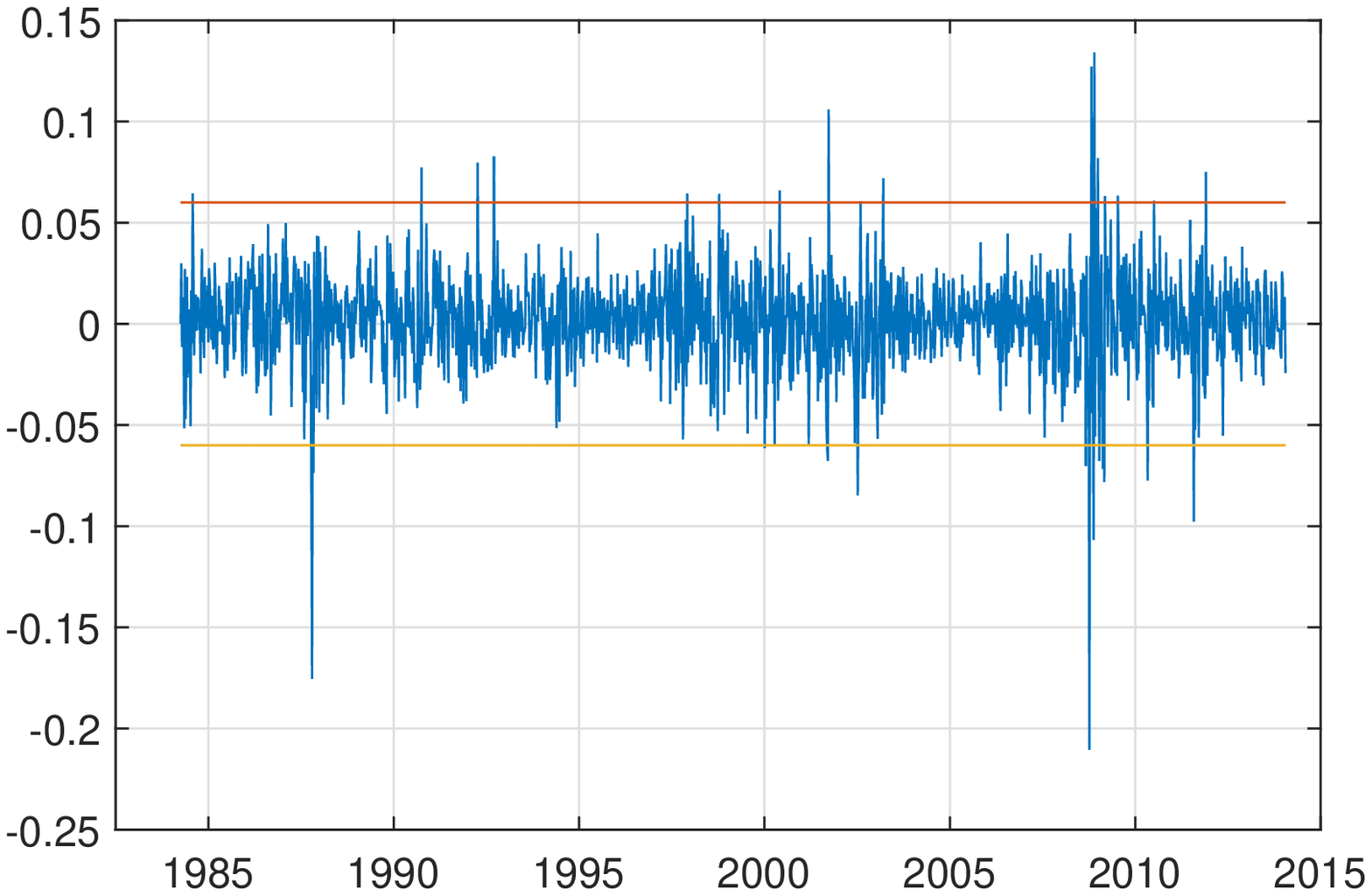} 

\vspace{-5mm}
\begin{center}(b) Returns.\end{center}
\end{minipage}
\caption{FTSE100 index: weekly values 1994-2013}
\label{ftse}
\end{figure}

As a representative data set we will take the series displayed in Figure \ref{ftse}(a), 30 years of weekly values $S_n$ of the FTSE100 stock index 1984-2013.
The accompanying Figure \ref{ftse}(b) shows the associated series of returns $Y_n=(S_n-S_{n-1})/S_{n-1}$ and demonstrates the typical stylised features found in financial price data: apparent non-stationarity and highly `bursty' volatility. The horizontal rules are at levels $\pm 0.06$, approximately the 1\% and 99\% quantiles. These will be needed in Section \ref{ru-text}.
\begin{figure}[h]
\centering
\begin{minipage}[b]{0.45\linewidth}
\includegraphics[scale=0.46]{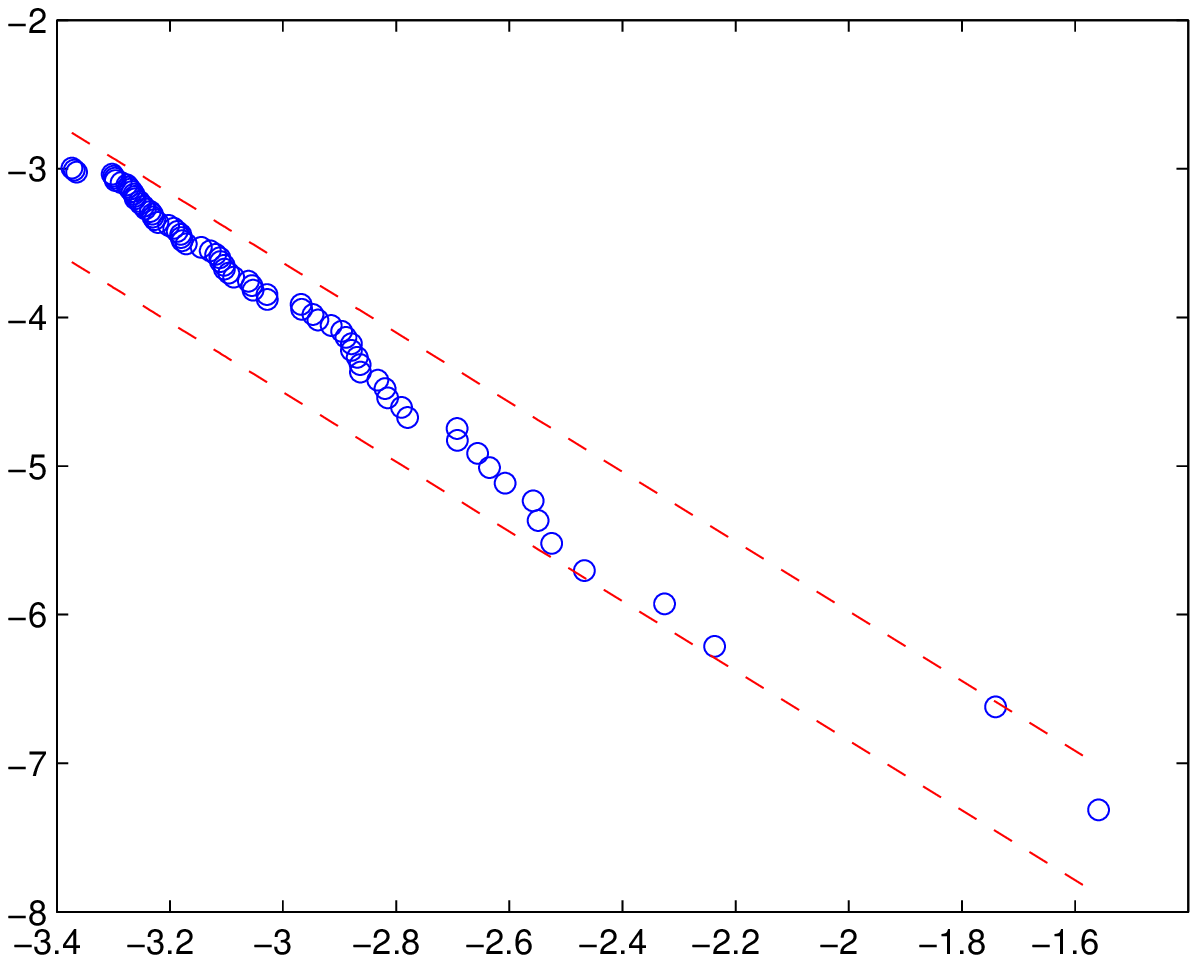}

\vspace{-5mm}
\begin{center}(a) Left tail ($x$-axis reversed)\end{center}
\end{minipage}
\qquad
\begin{minipage}[b]{0.45\linewidth}
\includegraphics[scale=0.46]{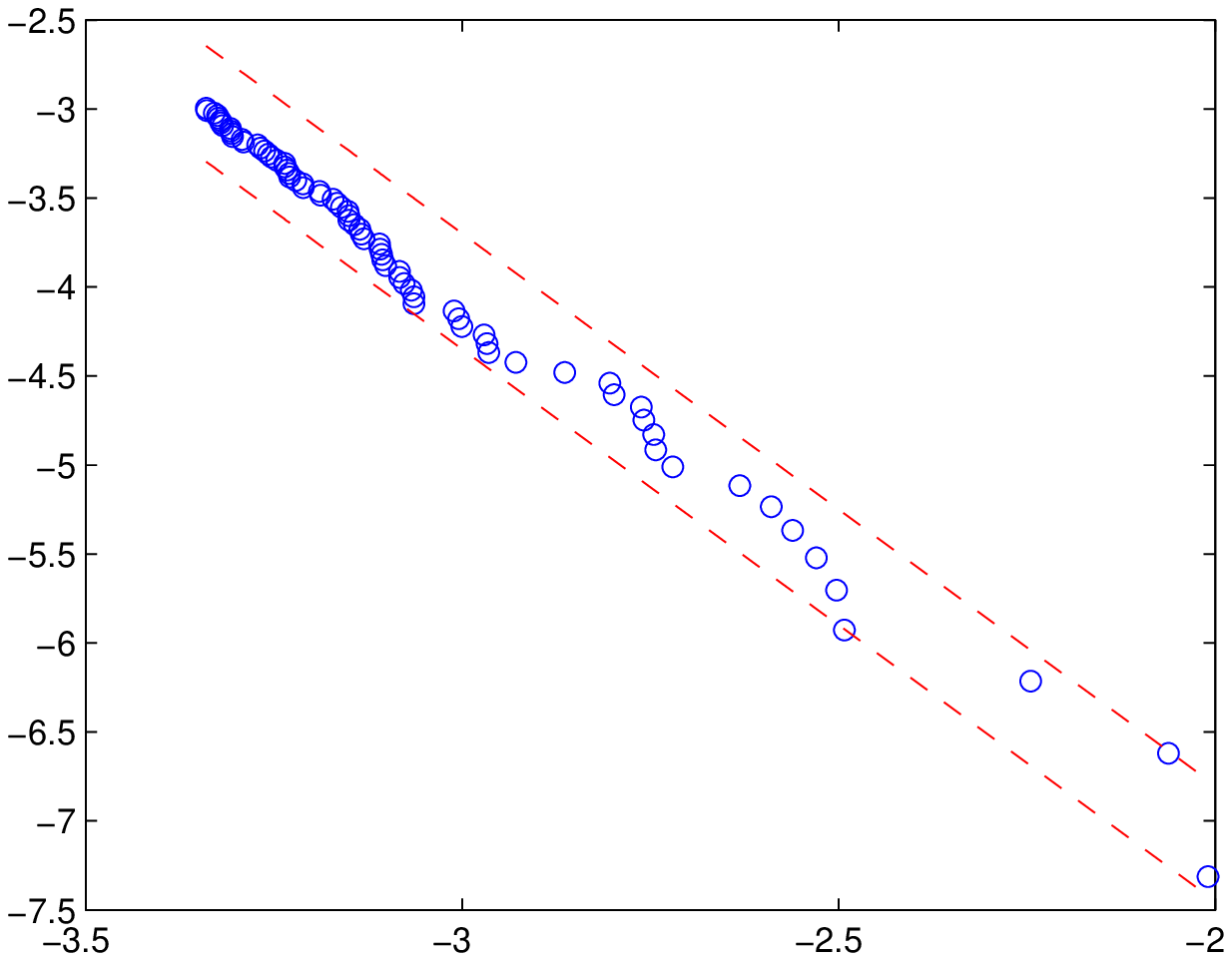}

\vspace{-5mm}
\begin{center}(b) Right tail\end{center}
\end{minipage}
\caption{Empirical distribution, left and right 5\% on log-log scale.}
\label{ed}
\end{figure}

Figure \ref{ed} shows the left and right 5\% =50 points of the empirical return distribution on a log-log scale. Based on this data, we conclude that the empirical distribution has power law tails with indices $\kappa=2.35, 3.25$ respectively
\footnote{A distribution function $F$ on $\bR$ has power left
tail with index $\kappa>0$ if $F(x)\sim|x|^{-\kappa}$ as $x\to-\infty$. To estimate $\kappa$ from an empirical tail sample we find, for a given $\kappa$, the tightest $0<c_1<c_2$ such that $c_1|x|^{-\kappa}\leq F(x)\leq c_2|x|^{-\kappa}$ for all $x$ in the sample and then minimise $c_2/c_1$ over $\kappa$, giving the bounds shown in Figure \ref{ed}. An analogous procedure applies to the right tail.}.
 Of course, these series have been the subject of intensive research over at least the last 50 years. A notable---and perhaps the most original---contributor was Beno\^\i t Mandelbrot \citep{mantay67, man97}, who introduced the heavy-tailed \emph{fractional Brownian motion} as an asset price model,  and the subject has become mainstream in econometrics \citep[see for example][]{bol86,clm90} and statistics \citep*{chbn96}. An excellent account from a `quant' perspective is \citet{con01}. It is however a curious fact that remarkably little of this effort has been aimed directly at prediction\footnote{There is an elegant paper on prediction of fractional Brownian motion by \citet{grinor96}.}, although from the perspective of contemporary risk management little else matters.

\subsection{External risk measures}\label{erm}
We mentioned above the distinction drawn by \citet{kph13} between `external' and `internal' risk measures. External risk measures are those employed by industry regulators in procedures imposed across the board on all regulated institutions, while internal risk measures are those used in individual institutions, or even trading desks, for day-to-day monitoring of the risks of trading books.

\begin{center}
\begin{figure}[h]
\includegraphics[scale=0.5]{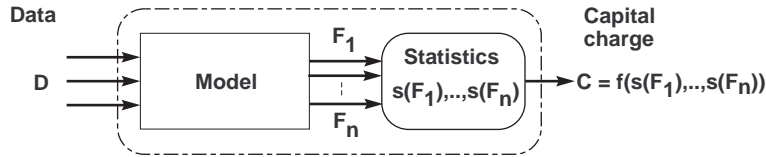}
\caption{Capital charge allocation process.}
\label{external}
\end{figure} 
\end{center}

The external/internal distinction mirrors exactly the distinction between cases (b) and (a) respectively of weather forecasting, discussed in Section \ref{sec:fund}(iii) above. External risk measures are part of the process, depicted in Figure \ref{external}, by which the regulator imposes capital charges on the bank in order to provide an adequate cushion against trading losses \citep{bis13}. Models must be built to compute return distributions $F_k,\, k=1\clc n,$ under various scenarios, and then the capital charge $C$ is equal to some function of $\{\mathfrak{s}(F_1)\clc \mathfrak{s}(F_n)\}$, where $\mathfrak{s}$ is the VaR or CVaR at some very high level such as 99.5\% or 99.75\%. The input data $D$ may be taken from the immediate past or/and from `stressed' periods in history, and is used for model calibration. Having calibrated the model, computations are invariably done by Monte Carlo simulation, so the computed $F_k$ \emph{always} have finite support; this point is stressed by \citet{kph13} and by \citet{rocury02} and is relevant to our discussion of elicitability below. There is no conceptual issue relating to the meaning of the return distributions $F_k$ since they are outputs from a well-defined stochastic model.
The whole process is an engineering job exactly analogous to flood barrier design.  The important question in this area is numerical stability: we want to avoid a situation where widely differing capital charges are imposed on different banks merely because of minor variations in their internal models; see \citet*{cds10} for an excellent study of this problem. A much bigger question, of course, is whether the decomposition implied by Figure \ref{external} of the map $D\mapsto C$ is the best way to arrive at an appropriate capital cushion. There are dissenting voices, see for example \cite{hal12}. 
In this paper we focus on \emph{internal} risk management---predicting the risks faced by trading books. Here the quantile level is typically much lower, say 95\%, so we can expect to see occasional exceedances and can monitor their frequency. This is the kind of problem probability forecasting is designed to address. 

\subsection{Falsifiability}\label{popper} First, let us consider a foundational issue. We compute what we claim to be the conditional distribution $F$ of future returns and/or some statistic $\fs(F)$.  But in this universe of highly non-stationary data, and given that no resampling is possible, we might well ask whether the predictive distributions implied by statistical models have any meaning at all.  A useful reference point is the falsifiability test of Karl \citet{pop02}: a statement is \emph{meaningful} if and only if it is \emph{falsifiable}, i.e. evidence could in principle be produced that would show the statement to be false\footnote{The concept is related to the basic asymmetry between proof and counterexample: to show $A\Rightarrow B$ we have to show that in \emph{every} case where $A$ holds, $B$ holds too, whereas to show 
$A\nRightarrow B$ we only have to find \emph{one} case where $A$ holds but $B$ does not.}. Now consider the following statement $\mathfrak{S}$: `the conditional distribution of the FTSE100 return $Y_k$, given data up to time $k-1$, is $F$', where $F$ is a specified distribution function. According to Popper's criterion, statement $\mathfrak{S}$ is surely meaningless. We compute $F$ at time $k-1$, and at time $n$ we get a single number $Y_k=x$; so was $F$ correct?  $\mathfrak{S}$ is falsified at time $k$ only if $x$ lies outside the support of $F$, which will never be the case in practice, where the support is invariably specified as $\bR$ (or $\bR^+$ for long-only portfolios). Since subsequent data points $Y_{k+1}, Y_{k+2},\ldots$ are drawn from different conditional distributions, they cannot be said to provide much useful evidence about the correctness of $F$, and in any case \emph{post hoc} data is not germane, since decisions have to be made on the basis of calculations at time $k-1$ and history cannot be rewritten afterwards. Consequently $\mathfrak{S}$ is not falsifiable, meaning that any statement about $F$ must depend on uncheckable \emph{a priori} modelling assumptions.

What is needed here is a shift of perspective. Instead of asking whether our model is correct, we should ask whether our objective in building the model has been achieved. This view of the prediction problem is standard in in some other areas of science, and was in fact pioneered in connection with weather forecasting \citep[see for example][]{js03}. An example will illustrate the point, taken from \cite{daw86}.
In many countries it is customary for weather forecasters to predict the probability of rain the next day in quantised form $0.0, 0.1, 0.2,\ldots,1.0$. The obvious way to evaluate such forecasts is, for each $n=0,\ldots,10$, to calculate over time the relative frequency of rain on days after the forecast probability was $n/10$. We then plot the relative frequencies against the predicted probabilities to obtain a \emph{reliability diagram}. Nothing is or can be asserted about the accuracy of the forecast on \emph{any particular} day. 
\begin{center}
\begin{figure}[h]
\includegraphics[scale=0.35]{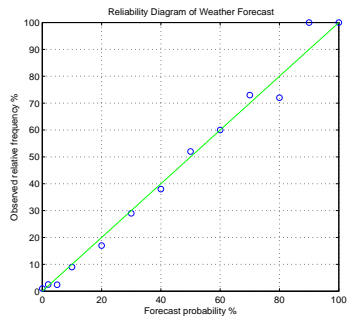}
\caption{Reliability diagram for Chicago forecaster.}
\label{chicago}
\end{figure} 
\end{center}
Figure \ref{chicago} shows the results of 2820 12-hour forecasts  by a forecaster in Chicago in the period 1972-76 \citep{daw86}; no-one can be in any doubt that this forecaster was doing a good job: the forecasts are well-calibrated. A key point here is  that the evaluation accords with the principles of P.Dawid's `prequential' theory of statistics \citep{daw84}. These principles, as enunciated in \citet{dawvov99} are

\smallskip\subhead{Weak prequential principle:} Evaluation of forecasting systems should be based only on the observed data and the numerical values of the forecasts produced (not on the algorithm that produced them).

\smallskip\subhead{Strong prequential principle:} Criteria for correct prediction should only depend on agreement between Nature and Forecaster on the stochastic law $\bP$ generating the data, not on what that law is (within some specified class $\cP$).

\smallskip Although formal application of these principles barely figures in the published literature on risk management, related methods are universally applied in the industry under the name of `back-testing'\footnote{This term is perhaps misleading as it seems to imply some special programme to re-live history rather than a procedure that is part of day-to-day practice. `Monitoring' would be a better description.}  A large part of the `VaR vs. CVaR' debate is concerned with the question whether it is true that if a statistic is not elicitable then it cannot be back-tested, see \citet{acsz14} for a recent contribution.  While we do not settle this question here, we do provide some formal structure within which the question can formulated in more precise terms.

\section{Elicitability}\label{sec:el}
The initial motivation for writing this paper was the striking set of results obtained by  \citet{gne11} and \citet{zie14} on the elicitability properties of VaR and CVaR. These authors showed that CVaR is not elicitable, and this was used in various quarters as an argument against its use as a risk-management statistic in place of VaR, which is elicitable. This argument has now fallen by the wayside, as \citet{fz15} have recently shown that the pair (VaR, CVaR) is \emph{jointly} elicitable, but nonetheless the controversy brought something new and important to the world of risk management. 
The circle of ideas relates to a decision-theoretic framework whose origins go back at least to work by L.J. \citet{sav71}, but the elicitability concept itself is due to \citet{osbrei85} and the name was coined by \citet{lps08}. The reader can consult \citet{gne11} for a wide-ranging exposition of this subject.

We consider the probability space $(\bR,\cB,\bP)$, where $\cB$ is the Borel $\sigma$-field. $Y$ will denote the identity function $Y(y)=y\in\bR$ and as usual the probability measure $\bP$ is identified with the (right-continuous) distribution function of $Y$. It is a familiar fact that if $Y\in L_2(\bR,\cB,\bP)$ then the function $f(x)=\bE[(x-Y)^2]$ achieves its minimum at $x=\bE[Y]$ and this is true whatever the distribution $F$ within the $L_2$ class. Elicitability is concerned with generalizing this characterization of the mean value to other statistics $\fs(F)$ of the distribution function. For a given statistic $\fs(F)$, can we find a \emph{score function} $S(x,y)$ such that $x\mapsto\bE_F[S(x,Y)]=\int S(x,y)F(dy)$ is minimized at $x=\fs(F)$ for all $F$ in some wide class $\cF$ of distributions? In general $\fs(F)$ may be set-valued, as is the case for the $\beta$-\emph{quantile}, and \emph{a fortiori} for the median, equal to $q_\ha$.

Our choice of score function will, as in \citet{gne11}, be restricted to  measurable functions $S:\bR^2\to\bR$ satisfying

(i) $S(x,y)\geq 0$ with equality if $x=y$

(ii) For each $y\in\bR$ the function $x\mapsto S(x,y)$ is continuous, and is continuously differentiable if $x\not= y$.

\noindent We say that $S$ is a \emph{consistent scoring function for a statistic $\fs$} relative to a class $\cF$ of distribution functions $F$ if whenever $Y\sim F\in\cF$
\be \bE_F[S(t,Y)]\leq\bE_F[S(x,Y)]\qquad\forall\,\, t\in\fs(F), x\in\bR.\label{cons}\ee
$S$ is \emph{strictly consistent} if it is consistent and equality in \eqref{cons} implies $x\in\fs(F)$.

\begin{definition} A statistic $\fs$ is \emph{elicitable for} $\cF$ if there exists a strictly consistent scoring function $S$.
\end{definition}

 The attractive feature of the approach is the precision of the results: it is possible to prove mathematically, in relevant cases, that particular statistics \emph{are} or \emph{are not} elicitable. What is not so clear is how to apply these results in a dynamic context such as risk management where the data is a sequence $Y_1, Y_2,\ldots$ of random variables each having a different conditional distribution, particularly in view of the fact that the criterion
\eqref{cons} fails to respect the weak prequential principle. The next section suggests an answer to this question, but first we consider a few examples, following \citet[][\S3]{gne11}.

\subsection{Examples}
\subsubsection{Mean value}  Here $\cF$ is the set of distributions with finite variance and the score function $S(x,y)=(x-y)^2$ is continuously differentiable. We can characterize optimality by noting that
\be\frac{\d}{\d x}\bE[S(x,Y)]=\bE\left[\frac{\d}{\d x}S(x,Y)\right]=x-\bE[Y],\label{dsdx-m}\ee
confirming that the expected score is indeed minimized at the mean value $\bE[Y]$. $S=(x-y)^2$ is not the only score function eliciting the mean value---others exist that do not require the existence of second moments; see Section \ref{expectile} below and \citet[][\S3]{gne11} for details  and further examples. 

\subsubsection{Quantiles and VaR}\label{sss-q} Here $\cF$ is the set of all probability distributions on some interval $I\subset\bR$. Then the $\beta$-quantile, $\beta\in(0,1)$ is elicitable. If $I$ is compact then a score function $S$ satisfying conditions (i), (ii) above is strictly consistent for the $\beta$-quantile if and only if it takes the form
\be S(x,y)=(\1_{(x\geq y)}-\beta)(g(x)-g(y))\label{Sq}\ee
where $g$ is a strictly increasing function. Score functions $S$ as in \eqref{Sq} are strictly consistent without the compactness assumption in the class of distributions for which the random variable $g(Y)$ is integrable. An obvious choice is $g(y)=y$, but if we take $g$ to be bounded and strictly increasing then no integrability condition is required.

Suppose $g$ is continuously differentiable and let $\cF_c$ be the class of continuous distribution functions. Then $S$ is continuously differentiable except at $x=y$ and
\be \frac{\d S}{\d x}=g'(x)[\1_{(x\geq y)}-\beta].\label{dsdx-q}\ee
Since the event $(Y=x)$ has probability 0 for all $F\in\cF_c$ we see that
\be  \bE\left[\frac{\d}{\d x}S(x,Y)\right]=g'(x)[F(x)-\beta],\label{ide}\ee
which is equal to zero if and only if $x$ is in the $\beta$-quantile set. If we drop continuity of the distribution function then
\be \frac{d}{dx}\bE[S(x,Y)]=g'(x)[(1-\beta)F(x-)+\beta F(x^+)-\beta].\label{dd}\ee
The expression on the right is negative if $x<q^-_\beta$ and positive if $x>q^+_\beta$, confirming that $\bE[S(x,Y)]$ is minimized at any $x$ in the $\beta$-quantile.

The value at risk $\mathrm{VaR}_\beta$ picks out one element, $q_\beta^-$, from the quantile set. Consequently, VaR is elicitable only in the set $\cF_\uparrow\subset\cF$ of \emph{strictly increasing} distribution functions, for which the quantile set is a singleton.

\subsubsection{Expectiles}\label{expectile} For $\tau\in(0,1)$ and $F\in L_1$ the $\tau$-expectile is the unique solution $m_\tau$ to the equation
\[ \tau\int_{(x,\infty)}(y-x)F(dy)=(1-\tau)\int_{(-\infty,x)}(x-y)F(dy).\]
If $\phi$ is a $C^1$ strictly convex function, the score function
\[ S(x,y)=(\tau\1_{(x<y)}+(1-\tau)\1_{(x\geq y)})(\phi(y)-\phi(x)-\phi'(x)(y-x))\]
is strictly consistent for the $\tau$-expectile in the class of $F$ such that $Y$ and $\phi(Y)$ are $F$-integrable. The natural choice is $\phi(x)=x^2$ when $(\phi(y)-\phi(x)-\phi'(x)(y-x))=(y-x)^2$. If $\phi\in C^2$ then
\be \frac{\d S}{\d x}=\phi''(x)[\tau\1_{(x<y)}+(1-\tau)\1_{(x\geq y)}](x-y),\label{dsdx-e}\ee
and hence
\[ \bE\left[\frac{\d}{\d x}S(x,Y)\right]=-\phi''(x)\left[\tau\int_{(x,\infty)}(y-x)F(dy)-(1-\tau)\int_{(-\infty,x)}(x-y)F(dy)\right]\]
so that $\bE[(\d S/\d x)(S(x,Y))]=0\,\Leftrightarrow\,x=m_\tau$. This characterization only requires $Y$ to be $F$-integrable. Note that the mean is the $\ha$-expectile, so a by-product here is a range of possible alternative score functions for the mean.

\subsection{Identifiability}\label{ident} Given a class of distributions $\cF$, an \emph{identification function} for a statistic $\fs$ is a measurable function $V:\bR^2\to\bR$ such that the expectation $\bE[V(x,Y)]$ is well-defined whenever $Y\sim F\in\cF$ and $x\in\bR$, and 
\[ \bE_F[V(x,Y)]=0\quad \Leftrightarrow\quad x\in\fs(F).\]
A statistic $\fs$ is $\cF$-\emph{identifiable} if an identification function exists. It is clear from \eqref{ide} above that if $\fs$ is elicitable with score function $S$ then under sufficient regularity conditions $V(x,y)=\d S/\d x(x,y)$ is an identification function. There is a kind of converse to this result, known as `Osband's principle' \citep[][\S 2.4]{gne11} according to which score functions can be obtained from identification functions by a natural integration procedure. The relationship between the two is examined in detail by \citet{spwz14}; their Corollary 9 asserts that for a scalar, single-valued statistic, under certain conditions elicitability is equivalent to existence of a bounded identification function, but the conditions include existence of a dominating measure, which is generally too restrictive.

\subsection{Distributional forecasts}\label{distfor} One way to predict a statistic is to predict the whole distribution and then calculate the statistic of the predicted distribution. This is the situation when one builds a stochastic model for the process of interest: past data is used to estimate model parameters and then it is a purely computational problem, tackled by analytic methods or by simulation, to evaluate the predicted distribution at some time in the future. Distributional forecasts may be evaluated by the use of \emph{proper scoring rules} \citep[][\S2]{gnr07}. If $\mP$ denotes the set of probability measures, or equivalently the set of distribution functions, on $(\bR,\cB)$ then we can define $(\mP,\cB_\mP)$ as the Borel space corresponding to the topology of weak convergence.  A \emph{forecast} is a choice of $F\in\cF\subset\cB_\mP$ where $\cF$ is some designated set of distributions. A scoring rule is a measurable function $\tilde{S}:\cF\times\bR\to\bR$ such that the function $y\mapsto\tilde{S}(F,y)$ is $G$-integrable for all $G\in\cF$. We define $\mathbf{S}(F,G)=\int\tilde{S}(F,y)G(dy)$. Here $F$ is the prediction and $G$ the `true' distribution. The scoring rule\footnote{\citet{gnr07} maximise instead of minimising and allow extended real valued scoring rules.} is \emph{proper} if $\mathbf{S}(G,G)\leq\mathbf{S}(F,G)$ for all $F\in\cF$ and \emph{strictly proper} if $\mathbf{S}(F,G)=\mathbf{S}(G,G)$ implies $F=G$. If $S(x,y)$ is a score function for elicitation of a statistic $\fs$ then $\tilde{S}(F,y)=S(\fs(F),y)$ is a scoring rule, but strict consistency of $S$ does not imply strict properness of $\tilde{S}$.

One advantage of constructing distributional forecasts is the availability of the probability integral transform (PIT) as a diagnostic tool, which has been widely used in statistics and econometrics \citep[see for example][]{daw84,
dgt98,chr98,gbr07,mw11}. PIT refers to the basic fact that if a random variable $Y$ has continuous distribution function $F$ then the random variable $U=F(Y)$ has uniform $[0,1]$ distribution. Thus uniformity of samples of $Y$ suggests that the distribution $F$ has been correctly evaluated. We make extensive use of the PIT in a somewhat different context in Section \ref{sec:qf} below.

There is one class of problems in which `point forecasts' and `distributional forecasts' coalesce, namely the problem of predicting the success probability in Bernoulli trials, where of course the success probability \emph{is} the distribution. There is an extensive literature on this problem---see \citet{lsg11} for a recent example where martingale theory is used in a somewhat similar way to Section \ref{sec:mv} below.

\subsection{Dynamic Models}
Suppose we observe not just a single variate $Y$ but a sequence $Y_1,Y_2,\ldots$. Any corresponding stochastic model is then a discrete-time process on some probability space $(\Omega,\cG,\bP)$, for which we denote by $F_k(y)$ the conditional distribution of $Y_k$ given $Y_1\clc Y_{k-1}$:
\[ F_k(y)=\bP[Y_k\leq y|Y_1\clc Y_{k-1}].\]
Suppose that, for some class $\cF$ of distributions and for all sequences $y_1,y_2,\ldots$ 

\smallskip\noindent(i) $F_k(\cdot\,;y_1\clc y_{k-1})\in\cF$;

\noindent(ii) For a given statistic $\fs$ there is an identification function $V$ such that for $F\in\cF$\[x\in\fs(F)\Leftrightarrow\bE_F[V(x,Y)]=0.\] 

\smallskip\noindent Then when $x_k=\fs(F_k)$ we have \[\bE[V(x_k,Y_k)|Y_1\clc Y_{k-1}]=0,\] i.e. $B_j\stackrel{\Delta}{=}\sum_{k=1}^jV(x_k,Y_k)$ is a martingale. This is the basis of our approach in the remainder of the paper. Under the conditions stated by \citet{spwz14}, item (ii) is tantamount to stating that the statistic $\fs$ is elicitable in the class $\cF$.

\section{Calibration of  Predictions}\label{sec:cp}
Now let us return to the world of risk management, a dynamic situation in which, at time $k-1$ we have observed a real-valued price series $Y_1,\ldots,Y_{k-1}$ and an $\bR^r$-valued series of other data $H_1,\ldots,H_{k-1}$ and wish to make some prediction relating to the behaviour of $Y_k$.

A \emph{model} for the data is a discrete-time stochastic process $(\tY_k,\tH_k)$ defined on a stochastic basis $(\Omega, \cG, (\cG_k),\mathbb{P})$. We always take $(\Omega, \cG, (\cG_k))$ to be the canonical space for an $\bR^{1+r}$-valued process, i.e. $\Omega=\prod_{k=1}^\infty \bR^{1+r}_{(k)}$ (where each $\bR^{1+r}_{(k)}$ is a copy of $\bR^{1+r}$) equipped with the $\sig$-field $\cG$, the product $\sig$-field generated by the Borel $\sig$-field in each factor. For $\omega\in\Omega$ we write $\omega=(\omega_1,\omega_2,\ldots)
\equiv((\tY_1(\omega),\tH_1(\omega)),(\tY_2(\omega),\tH_2(\omega)),\ldots)$. The filtration $(\cG_k, k\geq 0)$ is then the natural filtration of the process
$(\tY_k,\tH_k)$, with $\cG_0=(\Omega,\emptyset)$. With this set-up, different models amount to different choices of the probability measure $\bP$ on the same probability space $\Omega$. Below we will consider families $\cP$ of probability measures, and we will use the notation $\cP=\{\bP^m, m\in\mathfrak{M}\}$, where $\mathfrak{M}$ is an arbitrary index set, to identify different elements $\bP^m$ of $\cP$. The expectation with respect to $\bP^m$ is denoted $\bE^m$. It is not assumed that the observed data is a sample function of some model $m\in\mathfrak{M}$, or indeed of any model.

\begin{lemma} Let\, $\bP^m$ be any probability measure on $(\Omega, \cG, (\cG_k))$ as defined above. Then for each $k\geq 1$ there is a regular right-continuous conditional distribution of $\tY_k$ given $\cG_{k-1}$, i.e. a function $F_k^m:\bR\times\Omega\to[0,1]$ such that (i) for a.e. $\omega$, $F_k^m(\cdot,\omega)$ is a distribution function on $\bR$ and (ii) for each $x\in \bR$, 
\[F^m_k(x,\omega)=\bP^m[Y_k\leq x|\cG_{k-1}]\quad \mathrm{ a.s. }\, (\bP^m).\]
\end{lemma}
\proof For $k=1$, $F_1^m(x,\omega)=\bP^m[\tY_1\leq x]$, the unconditional distribution function. For $k>1$ the assertions of the lemma only involve the finite-dimensional vector r.v. 
\[((\tY_1,\tH_1)\clc(\tY_{k-1},\tH_{k-1}),\tY_k)\in\bR^{k(1+r)-r}.\]
Existence of a regular conditional distribution follows from Theorem 10.2.2 of \citet{dud89}.
\hfill$\square$
 
\medskip With these preliminaries in place, we now want to introduce the concept of \emph{calibration} for a statistic $\fs$ relative to a class of models $\cP$. Let $\mathfrak{I}(\cP)$ denote the set of strictly increasing predictable processes $(b_n)$ on $(\Omega, (\cG_k))$ such that $\lim_{n\to\infty}b_n=\infty$ a.s. $\forall \bP^m\in\cP$; in this context, `predictable' means that for each $k$, $b_k$ is $\cG_{k-1}$-measurable. Often, $b_k$ will actually be deterministic.
A \emph{calibration function} is a measurable function $\ell:\bR^2\to\bR$,  chosen so that 
 \be \bE^m[\ell(\tY_k,\fs(F^m_k))|\cG_{k-1}]=0\label{martdiff}\ee
  for all $\bP^m$ in some class $\cP$. Formally, this property is equivalent to saying that $\ell$ is an identification function as defined in Section \ref{ident}, but we use a different notation since here there is an extra ingredient $b\in\mathfrak{I}(\cP)$, so $\ell$ is just one component of the pair $(\ell,b)$. The norming sequence $b_n$ has no direct counterpart in elicitability theory. We will see below that in the case of statistics $\fs$ involving expectations it may be necessary to take random norming sequences and then the conditions for calibration become more complicated.
\begin{definition}\label{def:cons} A statistic $\fs$ is $(\ell,b)$\emph{-calibrated} in a set $\cP=\{\bP^m:m\in\mathfrak{M}\}$ of probability measures on $(\Omega, \cG)$, where $\ell$ is a calibration function and $b\in\mathfrak{I}(\cP)$, if
\be \lim_{n\to\infty} \frac{1}{b_n}\sum_{k=1}^n \ell(\tY_k,\fs(F^m_k))=0\quad \mathbb{P}^m\mathrm{-a.s.\,\, for\, each\,\, } m\in\mathfrak{M}.\label{crit}\ee
\end{definition} 

\noindent The criterion \eqref{crit} only depends on realized values of data and numerical values of predictions, in accordance with the weak prequential principle.

\medskip In practice we observe the data sequence $(Y_1, H_1\clc (Y_{k-1},H_{k-1})$ and produce an estimate $\pi(k)$, based on some algorithm, for what we claim to be $\fs(F_k)$.  The point of the calibration process is to check whether the `statistics' $\pi(n)$ we produce can reasonably be accepted as relating to some putative `conditional distribution'. Specifically, the quality of our predictions is gauged by calculating
\[ J_n(Y,\pi)=\frac{1}{b_n}\sum_{k=1}^n \ell(Y_k,\pi(k)).\]
Calibration is a `reality check': it says that if $(Y_i, H(i))$ were actually a sample function of some process and we did use the correct predictor $\pi(i)=\fs(F_i)$ then the loss $J_n$ will tend to zero for large $n$, and this will be true whatever the model generating $Y(i)$, within the class $\cP$, so a small value of $J_n$ is evidence that our prediction procedure is well-calibrated.  The evidence is strongest when $\cP$ is a huge class of distributions and $b_n$ is the slowest-diverging sequence that guarantees convergence in \eqref{crit} for all $\bP\in\cP$. Calibration is however only a necessary condition. We will see in Section \ref{nonsense} below that there can be `nonsense' predictors that survive the calibration tests while being almost unrelated to the data. This means that, to complete the picture, we need more tests to determine whether our predictions are related to the data according to clearly-stated criteria. An example in the case of VaR estimation will be found in Section \ref{ru_test}.

\section{Quantile forecasting}\label{sec:qf} Quantile forecasting is in a sense the `dual' of probability forecasting. In the weather forecasting problem described in Section \ref{popper} the \emph{event} (rain/no rain) is always the same and we forecast the \emph{probabilities} $p_n$, while in quantile forecasting the probability is fixed, $p_n=1-\beta$ where $\beta$ is the significance level, and the forecaster specifies the \emph{event} ($\mathrm{loss}\geq q_n$) by selecting $q_n$.
As in Section \ref{sec:cp} our set of models is
\[ (\Omega, \cG, (\cG_k),(\tY_k,\tH_k),\mathbb{P}^m),\quad \mathbb{P}^m\in\cP\]
where $\cP$ is some class of measures and $F_k^m(x,\omega)$ is the conditional distribution function of $\tY_k$ given $\cG_{k-1}$ under measure $\bP^m\in\cP$. Let $\mathfrak{P}$ be the set of all probability measures on $(\Omega, \cG)$, and define
\be \cP_c=\{\bP^m\in\mathfrak{P}: \forall k, F_k^m(\cdot\, , \omega)\in\cF_c\mbox{  for almost all }\omega\in\Omega\}.\label{p0}\ee
Here, $\cF_c$ is the set of continuous distribution functions. For risk management applications, the continuity restriction is of no significance; no risk management model would  predict positive probability for \emph{specific values} of future prices\footnote{Unless the model is based on Monte-Carlo generated empirical distributions, in which case some form of smoothing would be required.}. So $\cP_c$ is the biggest relevant subset of $\mathfrak{P}$.

  The following result is a slight extension of a well-known result generally credited to \citet{ros52};
 as mentioned in Section \ref{distfor} it is widely used in statistics and econometrics, and it is also used by \citet{he14} in much the same context as here. We give a statement and simple proof below to stress the fact that the result imposes absolutely no restriction on the stochastic basis or the joint distribution of the $\tY_k$ beyond the requirement that all conditional distributions be continuous.
    As before, $F^m_1$ denotes the unconditional distribution function of $\tY_1$.

\begin{proposition}\label{A} Suppose $\bP^m\in\cP_c$, defined by \eqref{p0} above. Then the random variables
$U_k=F_k^m(\tY_k)$, $k=1,2,\ldots$ are i.i.d. with uniform distribution $U[0,1]$.
\end{proposition}

\proof There are at most countably many intervals $I_k, k=\ldots -1,0,1,2,\ldots$ of positive length such that $F^m_1$ takes constant value $v_k$ on $I_k$ with $v_k<v_{k+1}$. For $x\notin\cI=\bigcup_kI_k$, $F^m_1$ is 1-1 and 
$ \bP^m[U_1\leq u_1]=\bP^m[\tY_1\leq (F_1^m)^{-1}(u_1)]=u_1$. Since $\cI$ has $F^m_1$ measure 0 we conclude that $U_1\sim U[0,1]$. Similarly, $U_k\sim U[0,1]$ for each $k>1$. Now suppose that $U_1\clc U_n$ are independent for some $n$. Then
\bstar \bP^m[U_i\leq u_i,\,i=1\clc n+1]&=&\bE^m\left[\left(\prod_{i=1}^{n}\1_{(U_i\leq u_i)}\right)\bP^m[U_{n+1}\leq u_{n+1}|\cG_n]\right]\\
&=& \bE^m\left[\left(\prod_{i=1}^{n}\1_{(U_i\leq u_i)}\right)\right]u_{n+1}=\prod_{i=1}^{n+1}u_i.
\estar
Thus all finite-dimensional distributions of $(U_i)$ are i.i.d. $U[0,1]$.\hfill$\square$

\subsection{Calibration of quantile estimates} For $\beta\in(0,1)$ let $q^m_k$ denote the $\beta$'th quantile of $F^m_k$, i.e. $q_k^m=\inf\{x:F^m_k(x)\geq\beta\}$. $q_k^m$ is an $\cG_{k-1}$-measurable random variable for each $k>0$. We use the calibration function $\ell$ defined at \eqref{dsdx-q} above.

\begin{theorem}\label{thm:quantest} If \,$\bP^m\in\cP_c$, defined at \eqref{p0}, then for any sequence $b_n\in\mathfrak{I}(\cP)$,
\be\frac{1}{b_n}\frac{1}{n^{1/2}(\log\log n)^{1/2}}\sum_{k=1}^n(\1_{(Y_k\leq q^m_k)}-\beta)\to 0 \quad\mathrm{a.s. \,} (\bP^n)\label{consq}\ee
Thus the quantile statistic $\fs(F)=q_\beta$ is $(l,b')$-calibrated for $\cP_c$  in accordance with Definition \ref{def:cons}, where $\ell(x,q)=\1_{(x\leq q)}-\beta$ and $b'_k=b_k(k\log\log k)^{1/2}$.
\end{theorem}

\proof By monotonicity of the distribution function, $(Y_k\leq q^m_k)\Leftrightarrow(U_k\leq F^m_k(q^m_k))\Leftrightarrow(U_k\leq\beta)$.
The result now follows from Proposition \ref{A} and by applying the Law of the Iterated Logarithm (LIL) \citep[Theorem 12.5.1]{dud89} to the sequence of random variables $Z_k=\1_{(U_k\leq\beta)}-\beta$, which are i.i.d with mean 0 and variance $\beta(1-\beta)$. Indeed, define
\[ \zeta(n)=\frac{1}{\sig(2n\log\log n)^{1/2}}\sum_{k=1}^nZ_k\]
where $\sig=\sqrt{\beta(1-\beta)}$. Then the LIL asserts that, almost surely, 
\[ \limsup_{n\to\infty}\zeta(n)=1,\qquad\liminf_{n\to\infty}\zeta(n)=-1.\]
 The convergence in \eqref{consq} follows.\hfill$\square$

\medskip Of course, if convergence holds in \eqref{consq} then it also holds if we replace the sequence $b$ by $b''$ such that $b''_n\geq b_n$ for all $n$. In particular, the conventional relative frequency measure
\be \frac{1}{n}\sum_{k=1}^n(\1_{(Y_k\leq q^m_k)}-\beta)\label{relfreq}\ee
converges under the same conditions; this also follows directly from the Strong Law of Large Numbers (SLLN) \citep[Theorem 8.3.5.]{dud89}; however, the LIL gives a stronger result.

The striking thing about Theorem \ref{thm:quantest} is that calibration of quantile forecasting is obtained under essentially \emph{no} conditions on the mechanism generating the data. As we shall see below, we cannot expect any such strong result in estimating other risk measures.

Theorem \ref{thm:quantest} is a `theoretical' result in that \eqref{consq} is a tail property, unaffected by any initial segment of the data. Nonetheless, it is practically relevant to compute the relative frequency \eqref{relfreq}. As we will show in Section \ref{sec:algo} below, doing so can provide convincing evidence that our prediction procedure is well calibrated, i.e. produces the right relative frequency of threshold exceedances, consistent with $q^m_k$ being the true $\beta$-quantile of $F^m_k$. For further evidence, we could examine by statistical test the other claim of Proposition \ref{A}, namely that the random variables $(U_k)$, and hence the binary variates $\1_{(Y_k\leq q^m_k)}$ are independent. We address this issue next.

\subsection{A test for serial dependence} \label{ss-sd}

 Given our prediction algorithm and the data return sequence $Y_k$ we generate a sequence $\ma=(a_0,a_1,\ldots)$ of binary r.v. $a_k=\1_{(Y_k\leq q^m_k)}$. The above tests give confidence that that $\ma$ is consistent with a model in which $\bP[a_k=1]=\beta$. We now want to test the first `i' in i.i.d., the null hypothesis being
\[\mathfrak{H}_0: \mbox{The $a_k$ are i.i.d. with $\bP[a_k=1]=\beta$}.\]
There are many tests that address this problem; some references were given in Section \ref{distfor}. An obvious recourse would be to use a non-parametric test such as  the Wald-Wolfowitz `runs' test \citep[][\S6.2]{gich10}. However, since we already know the marginal probability $\beta$, and since
it seem seems unlikely that $a_j$ and $a_k$ will fail to be essentially independent when $j\ll k$, it seems appropriate to use a test for `local' dependence. For this, 
a possible set of alternatives is\footnote{\citet{chr98} considers Markov chain alternatives but without the stationarity condition.}
\[ \mH_{\beta,\theta}: \begin{array}{l}\mbox{$\ma$ is a sample from a 2-state Markov chain}\\ \mbox{with stationary distribution $\bP[a_k=1]=\beta$}.\end{array}\]
Under $\mH_{\beta,\theta}$ the transition probabilities are
\bstar \bP[a_0=1]&=&\beta\\
\bP[a_k=1|a_{k-1}=0]&=&\theta\\
\bP[a_k=1|a_{k-1}=1]&=&\theta'.\estar
The stationary distribution is $\beta$ if
\bstar \beta=\bP[a_1=1]&=&\bP[a_1=1|a_0=0](1-\beta)+\bP[a_1=1|a_0=1]\beta\\
&=&\theta(1-\beta)+\theta'\beta.\estar
Thus $\theta$ and $\theta'$ are related, for given $\beta$,  by
\be \theta'=1-\frac{1-\beta}{\beta}\theta,\label{qprime}\ee
so $\mH_{\beta,\theta}$ is a 1-parameter family indexed by $\theta\in[0,1]$, when $\beta\geq\ha$. Assuming $\beta\geq\ha$ is no loss of generality since otherwise we can interchange the roles of `0' and `1'.
The i.i.d. case is $\theta=\theta'=\beta$. The log likelihood ratio $\mathrm{LLR^n_\theta}(\ma)=d\bP_{\beta,\theta}/d\bP_0$ is given by
\[ \mathrm{LLR^n_\theta}(\ma) =\mathrm{const} + n_1\log(1-\theta)+n_2\log(1-\theta f)+(n-n_1-n_2)\log(\theta),\]
where $f=(1-\beta)/\beta$ and $n_1, n_2$ are the numbers of 00, 11 pairs respectively in $\ma$. We denote $\bar{n}_i=n_i/n,\,i=1,2$.

\begin{theorem}\label{thm-qhat} Suppose $\beta\geq\ha$. Then

(i) The maximum likelihood estimate of $\theta$ is
\be\hat{\theta}_\beta(\ma)=\frac{1}{2f}\left(1-\bar{n}_2+f(1-\bar{n}_1)-\sqrt{(f-c_1)^2+4f(c_1-c_2)}\right)\label{qhat}\ee
where $c_1=1-f\bar{n}_1-\bar{n}_2,\,\,c_2=1-\bar{n}_1-\bar{n}_2$.

(ii) The estimator is consistent: under $\mH_{\beta,\theta}$, almost surely as $n\to\infty$
\bstar \bar{n}_1\to n_1^*&=&(1-\theta)(1-\beta)\\
\bar{n}_2\to n_2^*&=&\beta-(1-\beta)\theta,\estar
and $\hat{\theta}_\beta(n_1^*,n_2^*)=\theta$.
\end{theorem}
The proof of this result is given in Propositions \ref{pA1} and \ref{pA2} in Appendix \ref{sec:mcm} below.

The hypothesis $\mH_0$ that the $a_i$ are independent is equivalent to $\mH_{\beta,\theta}$ with $\theta=\beta$. In this case, $n_1^*=(1-\beta)^2, n_2^*=\beta^2$ and $\hat{\theta}_\beta(n_1^*,n_2^*)=\beta$. We can use the results of Theorem \ref{thm-qhat} to define a 2-sided test at significance level $\gamma$ in which $\mH_0$ is rejected if $\hat{\theta}(\bar{n}_1, \bar{n}_2)\notin[t_1, t_2]$ where the intervals $[0,t_1)$ and $(t_2,1]$ each have probability $\frac{\gamma}{2}$ under $\mH_0$. The endpoints $t_1, t_2$ are easily determined by simulation. Tables \ref{t1} and \ref{t2} give their values when $\beta=0.9, 0.95$ respectively, for four values of the significance level $\gamma$. 

An application of this test is given in Section \ref{sec:algo} below.

\begin{table}[tbh]
\begin{center}
\begin{tabular}{|c|cc|cc|cc|}\hline
$\gamma$&\multicolumn{2}{c|}{Data length 250}&\multicolumn{2}{c|}{Data length 500}&\multicolumn{2}{c|}{Data length 1000}\\ \hline
1\% & 0.7038& 1.0000  &0.7785 &1.0000& 0.8201  & 0.9672 \\
5\% & 0.7676& 1.0000  & 0.8103 &0.9758& 0.8418  & 0.9538 \\
10\% & 0.7926& 1.0000  & 0.8272 &0.9652& 0.8519  & 0.9450 \\
50\% & 0.8643& 0.9437  & 0.8728 &0.9281& 0.8823  & 0.9200 \\ \hline
\end{tabular}
\vspace{-3.5mm}\caption{Confidence intervals $t_1,t_2$ for estimator $\hat{\theta}_\beta,\, \beta=0.90$.} 
\label{t1}
\end{center}
\end{table}

\begin{table}[tbh]
\begin{center}
\begin{tabular}{|c|cc|cc|cc|}\hline
$\gamma$&\multicolumn{2}{c|}{Data length 250}&\multicolumn{2}{c|}{Data length 500}&\multicolumn{2}{c|}{Data length 1000}\\ \hline
1\% & 0.6080& 1.0000  &0.7854 &1.0000& 0.8516  & 1.0000 \\
5\% & 0.7600& 1.0000  & 0.8398 &1.0000& 0.8800 & 1.0000\\
10\% & 0.8012& 1.0000  & 0.8648 &1.0000& 0.8940  & 1.0000 \\
50\% & 0.9133& 1.0000  & 0.9249 &1.0000&0.9308  & 0.9732 \\ \hline
\end{tabular} 
\vspace{-3mm}
\caption{Confidence intervals  $t_1,t_2$ for estimator $\hat{\theta}_\beta, \,\beta=0.95$.}
\label{t2}
\end{center}
\end{table}
\subsection{A `nonsense' quantile predictor}\label{nonsense} Even if a quantile predictor passes both the calibration and independence tests it may still be seriously deficient. A striking example illustrating this for the 95\% quantile was given by \citet[][\S3.1]{he14}. The quantile predictor is set at some very high level $h$ on 95 out of every 100 dates, and at a very low level $l$ on the remaining 5 dates. Then the empirical exceedence frequency will be almost exactly 5\%, although the predictor is barely related to the data. A variant of this example, already given by \citet{em04}, would be to take an i.i.d. Bernoulli sequence $B_k$ with $\bP[B_k=1]=0.05$ and define our quantile predictor as $\hat{q}_k=lB_k+h(1-B_k)$. Then almost always $\1_{Y_k\geq\hat{q}_k}=B_k$, so this predictor will pass both the LIL test of Theorem \ref{thm:quantest} and the independence test of Section \ref{ss-sd}.

Examples of this sort pose a major challenge to verification of prediction. In the case of distributional prediction, as discussed in Section \ref{distfor}, \citet{gbr07} proposed a diagnostic approach based on `maximising the sharpness of the predicted distributions subject to calibration'. The idea of `sharpness' is that, given two distributions, the one with the minimum dispersion (as measured by an inter-quantile range, for example) should be preferred. While this may be appropriate in certain applications such as predicting macroeconomic variables, it is not without controversy, see \citet{mw11}. In any case, the principle is not applicable for point forecasts as considered in this paper. Reacting to the `nonsense' example, a reasonable criterion might be that predictor $A$ is preferred to predictor $B$, given that both survive calibration tests, if $A$ is more sensitive to the data than $B$. If prediction is based on a data vector $X=(X_1\clc X_n)$, where $X_n$ is the most recent data point, we could compute directional derivatives $\partial_ZA=\lim_{\epsilon\downarrow 0}(A(X+\epsilon Z)-A(X))/\epsilon$ and $\partial_ZB$ for a range of deterministic pertubation vectors $Z$ and prefer $A$ to $B$ if $A$ has the greater average derivative. Obvious choices of $Z$ could be $\1=(1\clc 1)$ or $Z_k=\alpha^{n-k}$ with $\alpha\in(0,1)$ if sensitivity to recent data is thought to be more important. With $Z=\1$ the nonsense predictor has sensitivity practically equal to zero, while the quantile predictor introduced in Section \ref{sec:algo} has sensitivity close to 1. Further investigation of these ideas is a subject for future research. A completely different diagnostic that successfully separates these two predictors is described in Section \ref{ru-text} below. One thing all investigators are agreed upon is that, while calibration is---in accordance with the weak prequential principle---a property of the data and predictors jointly, diagnostics beyond calibration are functions of the predictors \emph{only}.

\section{Risk Measures Involving Mean Values}\label{sec:mv} 
Risk measures such as CVaR involve integration with respect to the conditional distribution functions $F^m_k$. In this section we will consider the straight prediction problem of estimating the conditional means
\be \mu^m_k=\int_\bR xF^m_k(dx)\label{mu}.\ee
We must assume that the class of candidate models is at most
\[ \cP^1=\left\{\bP^m\in\mathfrak{P}: \forall k, \int_\bR |x|F^m_k(dx)<\infty\right\}.\]
In this context, continuity of the conditional distributions is not required, so $\cP^1$ is not a subset of $\cP_c$. In fact, this problem is general enough to include risk measures of the form $\int f(x)F^m_k(dx)$ for general functions $f$: we can simply define a new model class $(\tY', \tH')$ where $\tY'_k=f(Y_k)$ and $\tH'_k=(Y_k,H_k)$. Note that if $f$ is an option-like function such as $f(x)=(x-K)^+$ then $f(\tY_k)=0$ with positive probability for some measures $\bP^m$, so it is convenient that we do not require $\bP^m\in\cP_c$.
\subsection{Universality}
The first question to ask is whether we can get any `universal' result, similar to Theorem \ref{thm:quantest}, for estimating $\mu^m_k$, by using the i.i.d. sequence $U_k$ of Proposition \ref{A}. The answer appears to be no. What makes Theorem \ref{thm:quantest} work is the equality
\[ \1_{(\tY_k\leq q^n_k)}-\beta=\1_{(U_k\leq\beta)}-\beta,\]
so by transforming the variables we obtain the universal calibration function $l(u,\beta)=\1_{(u\leq\beta)}-\beta$.
In the case of expected value prediction the natural criterion is
\[ \frac{1}{n}\sum_{k=1}^n(\tY_k-\mu^m_k)\to 0.\]
Mapping the two variables in the $k$th term through the distribution function $F_k^m$ gives us a summand
\[ U_k-F^m_k(\mu^m_k). \]
This translates into a universal calibration function if and only if there is a constant $c$ such that
\be F^m_k(\mu^m_k)=c\quad\mbox{ a.s. for all}\,\, \bP^m,\label{symm}\ee
 meaning that $\mu^m_k$ coincides with a \emph{fixed quantile} $c$ of $F^m_k$. But if that is the case the problem reduces to quantile estimation and the results of Section \ref{sec:qf} apply. The only natural example of this is the situation where each distribution function $F^m_k$ is symmetric around its mean value, when \eqref{symm} holds with $c=\ha$. This is not a relevant class from the risk-management perspective, but see \citet{ar95} where the relations between quantiles and expectiles are examined in greater detail.

\subsection{Martingale analysis}
To proceed further, we need to make use of martingale properties. If we define
\be X_k=\tY_k-\mu^n_k,\qquad S_n=\sum_{k=1}^nX_k\label{YS}\ee
with $S_0=0$, then $S_n$ is a zero-mean $\bP^m$-martingale since $\bE^m[X_k|\cG_{k-1}]=0$. We want to determine calibration conditions by using the SLLN for martingales. In this subject, a key role is played by the \emph{Kronecker Lemma} of real analysis.
\begin{lemma} Let $x_n, b_n$ be sequences of numbers such that $b_n>0$, $b_n\uparrow\infty$, and let $u_n=\sum_{k=1}^nx_n/b_n$. If $u_n\to u_\infty$ for some finite $u_\infty$ then
\[\lim_{n\to\infty}\frac{1}{b_n}\sum_{k=1}^nx_k =0.\] 
\end{lemma}

The \emph{martingale convergence theorem} states that if $S_n$ is a zero-mean martingale on a filtered probability space and there is a constant $K$ such that $\bE|S(n)|\leq K$ for all $n$, then $S_n\to S(\infty)$ a.s. where $S(\infty)$ is a random variable such that $\bE|S_\infty|<\infty$.

Now let $X_k, S_k$ be as defined at \eqref{YS} above, and let $Z_k$ be a \emph{predictable} process, i.e. $Z_k$ is $\cG_{k-1}$-measurable, such that $Z_k>0$ and $Z_k\uparrow\infty$ a.s. Let $X^Z_k=X_k/Z_k$ and $S^Z(n)=\sum_1^nX^Z_k$.
Then $S^Z_n$ is a martingale\footnote{$S^Z$ is a stochastic integral, or `martingale transform' of $S$.}, since
\[\bE^m[X^Z_k|\cG_{k-1}]=\frac{1}{Z_k}\bE^m[X_k|\cG_{k-1}]=0.\]
If we can find $Z_k$ such that $\bE^m|S^Z(n)|<c_Z$ for some constant $c_Z$ then $S^Z$ converges a.s. and hence by the Kronecker lemma
\[ \frac{1}{Z(n)}S(n)= \frac{1}{Z(n)}\sum_{k=1}^n(\tY_k-\mu^n_k)\to 0 \quad\mathrm{a.s.}\]
We have shown
\begin{proposition}\label{useless} Under the above conditions, the statistic $\fs(F)=\int  x F(dx)$ is $(\ell,Z)$ calibrated in the class $\cP^1$, according to the Definition \eqref{def:cons}, where $\ell(x,\mu)=x-\mu$.
\end{proposition}
Note that the calibration function $\ell$ is the one derived from elicitability, see \eqref{dsdx-m}.
The above proposition is of course useless as it stands, because no systematic way to specify the norming process $Z_k$ has been provided. We can partially resolve this problem by moving to a setting of
\emph{square-integrable martingales} \citep[see][Chapter 12]{wil91}. If $S(n)\in L_2$ we define the `angle-brackets' process $\aS_n$ by 
\be \aS_n=\sum_{k=1}^n\bE[X^2_k|\cG_{k-1}].\label{angle}\ee
This is the increasing process component in the Doob decomposition of the submartingale $S^2(n)$. 
\begin{proposition}[\citealt{wil91}]\label{DW} If $S(n)$ is a square-integrable martingale then $S(n)/\aS_n\to 0$ on the set $\{\omega: \aS_\infty(\omega)=\infty\}$.
\end{proposition}
\noindent\textit{Proof}
Define the martingale $W(n)=\sum_{k\leq n}X_k/(1+\aS_k)$, for which
\bstar \bE[(W(n)-W(n-1))^2|\cG_{n-1}]&=&\frac{1}{(1+\aS_n)^2}(\aS_n-\aS_{n-1})\\
&\leq&\frac{1}{1+\aS_{n-1}}-\frac{1}{1+\aS_n} \quad\mathrm{a.s.}\estar
It follows that $\langle W\rangle_\infty\leq 1$. From \citet[Theorem 12.13]{wil91} this implies that $\lim_n W_n$ exists, and hence from the Kronecker lemma that $S(n)/\aS_n\to 0$ as long as $\aS_n\uparrow\infty$.\hfill$\square$ 

\medskip Proposition \ref{DW} shows that in the square-integrable case we can take $Z=\aS$ in Proposition \ref{useless}. However, we cannot use $\aS$ as it stands because it does not satisfy the weak prequential principle, which requires that the norming sequence be calculable using only observed data and numerical values of estimates. To achieve this, we follow a line of reasoning pursued by \citet{halhey80}, relating the predictable quadratic variation $\aS_n$ to the realized quadratic variation
\[ Q_n = \sum_{k=1}^n(S_k-S_{k-1})^2= \sum_{k=1}^nY^2_k.\]
As Hall and Heyde point out, the two random variables $Q_n$  and $\langle S\rangle_n$, defined at \eqref{angle}, have the same expectation, and we are interested in the ratio $Q_n/\aS_n$. To get the picture, consider the case where the $Y_k$ are i.i.d. with variance $\sigma^2$. Then $\aS_n=\sigma^2n$ and
\be\lim_{n\to\infty} \frac{Q_n}{\aS_n}=\frac{1}{\sigma^2}\lim_{n\to\infty}\frac{1}{n}
\sum_{k=1}^nY^2_k=1\quad\mathrm{a.s.}\label{Q}\ee
by the SLLN. In the general, martingale, case we may or may not have convergence as in \eqref{Q}, as \citet{halhey80} show. We will not present details of their analysis here but content ourselves with the following definition.
\begin{definition}\label{pe} Let $\cP^e\subset\mathfrak{P}$ be the set of probability measures $\bP^m$ such that

(i) $\forall k,\,\,\tY_k\in L_2(\bP^m)$.

(ii) $\lim_{n\to\infty}\aS_n=\infty\,\,\mathrm{a.s.}\,\bP^m$, where $S_n$ is defined at \eqref{YS}.

(iii) There exists $\epsilon_m>0$ such that $Q_n/\aS_n>\epsilon_m$ for large $n$, a.s. $\bP^m$.
\end{definition}

\noindent We can now state our final result.
\begin{theorem}\label{thm:meantest} The mean statistic $\fs(F)=\int xF(dx)$ is $(\ell,Q_n)$ calibrated for the class $\cP^e$, where \[\ell(x,\mu)=x-\mu.\]
\end{theorem}
\proof Suppose $\bP^m\in\cP^e$. Conditions (i) and (ii) of Definition \ref{pe} imply that $S(n)/\aS_n\to 0$ by Proposition \ref{DW}. Using condition (iii) we have
\[ \left|\frac{S(n)}{Q_n}\right|=\frac{\aS_n}{Q_n}\left|\frac{S(n)}{\aS_n}\right|
         \leq \frac{1}{\epsilon_m}\left|\frac{S(n)}{\aS_n}\right|\quad\mbox{for large $n$}.\]
The result follows.\hfill$\square$

\medskip As we see, significant conditions must be imposed to secure consistency of mean-type estimates, in contrast to the situation for quantile estimates (Theorem \ref{thm:quantest}) where almost \emph{no} conditions are imposed. Theorem \ref{thm:quantest} is a LIL-based result whereas Theorem \ref{thm:meantest} is based on the SLLN. There is a sizable literature on the LIL for martingales \citep[see][again]{halhey80}, but a number of quite intricate conditions are required, none of which would be checkable in the context of mean estimation, so it does not seem worth pursuing this avenue here.  

Arguments based on martingale convergence have been used in similar contexts by \citet{dawvov99} and \citet{lsg11}.

Verifying the validity of mean-based estimates is always more problematic than the same problem for quantile-based statistics. In fact the whole process of mean estimation is more problematic because, just from the basic definition \eqref{mu}, the mean depends in an essential way on the tail of the distribution function $F$ and, in any situation involving real data rather than model-generated data, we run out of data at some point in trying to estimate the tail, but the unestimated part may contribute significantly to the mean. We discuss in the next section question of CVaR estimation where this difficulty can be seen very clearly.


\section{Estimating CVaR}\label{sec:cvar} This section focuses on computation of CVaR. It has been pointed out by \citet*{cds10} that CVaR is excessively sensitive to small changes in the data sequence from which it is computed. Here we wish to make the more general point that any mean calculation depends on the tails of the distribution in ways that cannot be easily controlled. This is discussed in Section \ref{bp} below but first, in Section \ref{ru-text} we introduce the characterisation, due to \citet{rocury00}, of CVaR as the minimum of a certain convex function. Recent work
by \citet{he14} has highlighted the relation between this result and elicitability.

\subsection{CVaR as the solution to a minimisation problem}\label{ru-text} Let $F$ belong to the set $\cF_{c\uparrow}$ of continuous and strictly increasing distribution functions on $\bR^+$. From \eqref{cvar1}, the CVaR at level $\beta$ can be expressed as 
\[ \mathrm{CVaR}_\beta(F)=q_\beta+\frac{1}{1-\beta}\int_{q_\beta}^{\infty}(y-q_\beta)F(dy)
=\frac{1}{1-\beta}\int_\beta^1q_\tau d\tau.\]
where $q_\tau$ is the unique $\tau$-quantile of $F$. \citet{rocury00,rocury02} give a charaterisation of CVaR as the solution to a minimization problem given as follows. \begin{proposition} For $x\in\bR, \beta\in(0,1)$ let
\[ \Psi^F_\beta(x)=x+\frac{1}{1-\beta}\int_x^\infty(y-x)F(dx).\]
Then \[\mathrm{CVaR_\beta}(F)=\min_{x\in\bR}\Psi^F_\beta(x)=\Psi^F_\beta(q_\beta).\]
\end{proposition}
It has been noted by \citet[][\S3.2]{he14} that this result is closely related to the elicitability properties of the quantile $q_\beta$. Recall from Section \ref{sss-q} that score functions for the quantile take the form
$S(x,y)=(\1_{x\geq y}-\beta)(g(x)-g(y))$.
If we take $g(x)=x/(1-\beta)$ then we find that 
\be\label{S*} S(x,y)=x+\frac{1}{1-\beta}(y-x)\1_{y>x}-\frac{y}{1-\beta}\,\,\stackrel{\Delta}{=}\,\,
S^*(x,y)-\frac{y}{1-\beta}.\ee
The term $y/(1-\beta)$ plays no role in the minimization, and $\bE_F[S^*(x,Y)]=\Psi^F_\beta(x)$.
Thus the minimum value of $\bE_F[S^*(x,Y)]$, achieved at $x=q_\beta$, is exactly $\mathrm{CVaR}_\beta$.

\begin{figure}[h]
\centering\includegraphics[scale=0.4]{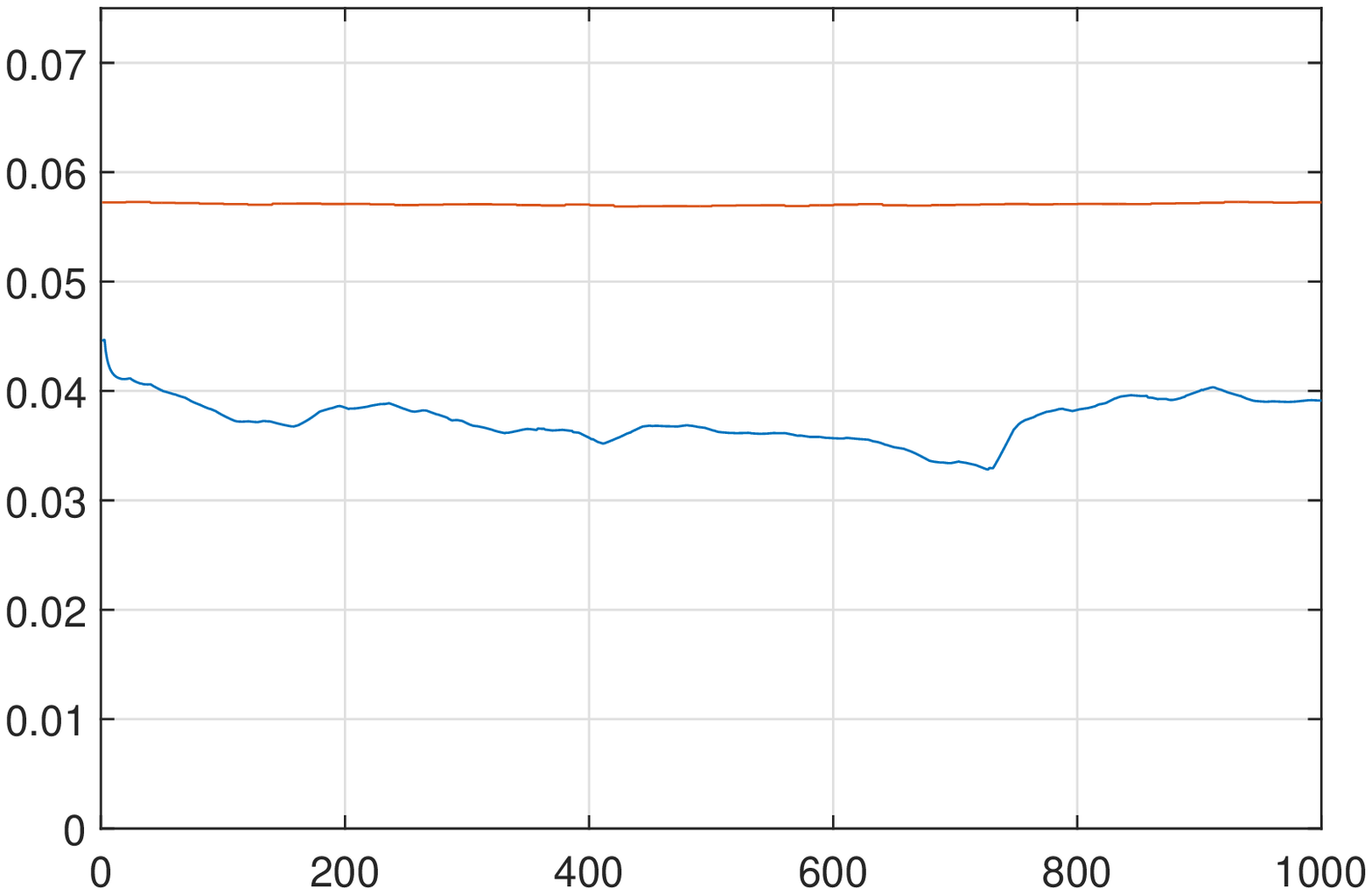}
\caption{Plot of $x^1_j$ and $x^2_j$, defined by \eqref{xmj}, against $j=1\clc1000$. Lower curve is $x^1_j.$ }
\label{ru_test}
\end{figure}

This result gives us a diagnostic test for comparing VaR estimators that have survived calibration and independence tests. Given a data sequence $Y_1,Y_2, \ldots$, let $\hat{q}_k^m, k=1,2\ldots$ be the sequences of $\beta$-quantile predictions produced by two algorithms $m=1,2$. Then we prefer predictor 1 to predictor 2 if
\be \frac{1}{n}\sum_{k=1}^n S^*(\hat{q}^1_k,Y_k)<\frac{1}{n}\sum_{k=1}^n S^*(\hat{q}^2_k,Y_k).\label{apples}\ee
This procedure of comparing realized average scores is known as a Diebolt-Mariano test \citep{dm95}, see also  \citet[][\S1.1]{gne11}. There is no rigorous justification for the test without extra mixing or other assumptions, but nonetheless it proves to be effective. As an example, let $\hat{q}^1_k$ be the predictors of 90\% return quantiles for the FTSE100 data of Figure \ref{ftse} produced by the algorithm \eqref{qp1}, \eqref{qp2} described in Section \ref{sec:algo} below, and let $\hat{q}^2_k$ be Holzmann and Eulert's `nonsense' predictor as described in Section \ref{nonsense} above. The two parameters $l,h$ for this algorithm are taken as $l=-0.06, h=+0.06$, which are close to being lower and upper bounds for the return sequence, see Figure \ref{ftse}(b). We compute the averages as in \eqref{apples} fixing $n=500$ but taking a moving window of data. Specifically, we compute
\be x^m_j=\frac{1}{500}\sum_{k=j}^{j+499} S^*(\hat{q}^m_k,Y_k), \quad j=1\clc1000,\,m=1,2.\label{xmj}\ee
As can be seen from Figure \ref{ru_test}, the algorithm of Section \ref{sec:algo} is consistently and decisively preferred to the nonsense algorithm.

\subsection{Basic problems in CVaR estimation}\label{bp}
In Section \ref{finrisk} we saw that the empirical distribution of returns for the FTSE100 data set displayed power tails (defined precisely in the footnote there) with tail index 2.35 on the left (=loss) side. We should not read too much into this since it is not claimed that the returns are samples from the same distribution, but nevertheless it does add credibility to the idea of considering power-tail distributions as candidates for a model in the sense defined in Section \ref{sec:cp}.

To clarify the difficulty in CVaR estimation, consider the following proposition, in which $F$ is supposed to have \emph{exact} power tail. Its proof is a simple computation.

\begin{proposition} Let $0<\beta<\eta<1$ and $F$ be a continuous distribution function on $\bR^+$ such that for $x\geq q_\eta^+$
\[ F(x)=1-(1-\eta)\left(\frac{x}{q_\eta}\right)^{-\kappa}\]  
where $\kappa>1$. Then
\be  \mathrm{CVaR}_\beta(F)=\frac{1}{1-\beta}\left(\int_\beta^\eta q_\tau d\tau +\frac{\kappa}{\kappa-1}(1-\eta)q_\eta\right).\label{cv}\ee
\end{proposition}

It will be seen in the next section that quantile estimation for financial data is something that can be achieved convincingly for significance levels out to $95\%$ at least. A further point is that the representation clearly relates
to the idea of estimating distributions by estimating a series of quantiles, see \citet{cm96} or \citet[][\S6.1]{gnr07} for further details.
Suppose we wish to compute $\mathrm{CVaR}_\beta$ and can reliably estimate quantiles $q_\tau$ for $\tau\leq\eta$ but not beyond $\eta$ where the data has dried up. Then the first term on the right of \eqref{cv} and the value of $q_\eta$ are known,  but the result also depends on the value of $\kappa$, and $\mathrm{CVaR}_\beta(F)\to+\infty$ as $\kappa\downarrow 1$. To place an upper bound on CVaR requires a reliable estimate for the tail index $\kappa$ but by definition this is impossible to obtain.  The conclusion is that any estimate of CVaR depends on \emph{a priori} assumptions about tail behaviour that cannot be verified on the basis of any finite data set, however large.

Various expedients can be employed:

\smallskip\ni(i) If the empirical return data exhibits power tails, for example the FTSE100 data where the left (=loss)  tail index is $\kappa=2.35$, then use this value beyond the last point where the quantiles can be accurately estimated. 

\smallskip\ni(ii) Use an econometric model. Any model implies tail behaviour, which might be time and data-dependent.

\smallskip\ni(iii) Use methods based on extreme-value theory \citep{ekm97}. 

\smallskip\ni(iv) Extrapolation: given reliable estimates for $q_\beta$ and $q_\eta$ and assuming one is already in the tail regime at $q_\beta$ one can back out the implied value of $\kappa$. This is however likely to be a very noisy estimate.

\smallskip\ni(v) \citet{cds10} suggest modifying the definition of $\mathrm{CVaR}_\beta$ to
\[ \frac{1}{\eta-\beta}\int_\beta^\eta q_\tau d\tau,\quad\mbox{for some}\,\,\eta<1,\]
providing a robustly computable statistic.

\smallskip\ni(vi) \citet{kph13} propose replacing CVaR by CMVaR, the conditional \emph{median} loss beyond VaR. Clearly, $\mathrm{CMVaR}_\beta=\mathrm{VaR}_{\eta}$ with $\eta=(1+\beta)/2$, so computation reduces to VaR estimation.

\smallskip All of these have their disadvantages. Item (i) assumes a relationship between the empirical distributions and the conditional distributions of CVaR which cannot be rigorously justified. Item (ii) is a broad-brush approach which typically uses empirical tail estimates to infer a suitable choice of i.i.d. drivers for the model.
In item (iii), extreme-value theory is an analysis of i.i.d. samples and in the present context is best suited to the very high significance levels $\beta$ of external risk management.
Item (iv) is predicated on power tail and is certainly not `robust', while the remaining  two cut off the tail completely at some point, which might miss real risk. Both of these require reliable VaR estimates up to level $\eta>\beta$ where $\beta$ is the level at which CVaR is required.

From a practical perspective the purpose of computing CVaR is to establish some threshold beyond VaR such that the gap $(\mathrm{CVaR}-\mathrm{VaR})$ provides an adequate cushion against extreme losses. From that point of view, CMVaR seems the right choice in that it is easy to compute, has a clear statistical meaning and has axiomatic support \citep{kph13}.

\section{An algorithm for quantile forecasting}\label{sec:algo}
By way of illustration, we present in this section a very simple data-driven algorithm for producing 1-week ahead forecasts of the upper 10\%  or 5\% quantiles of returns on the FTSE100 index. The data is weekly values of the index 1994-2013. Figure \ref{ftse} shows the index values and the series of weekly returns. This series exhibits the usual `stylized facts': gross variations in volatility and apparent non-stationarity.

\begin{center}
\begin{figure}[h!] 
\includegraphics[scale=0.5]{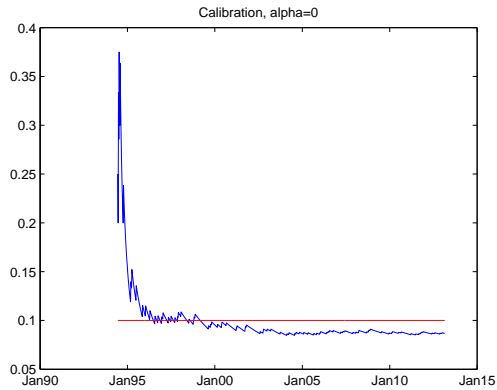} 
\caption{FTSE100 Calibration of initial algorithm.}
\label{cal0}
\end{figure}  
\end{center}

To predict quantiles of the returns the traditional approach is to estimate the parameters of an econometric model such as GARCH or EGARCH and then compute the 1-week ahead conditional distribution. However, a much simpler data-driven approach seems competitive in terms of calibration. As always, we are only checking necessary conditions with these tests. We start with the 90\% quantile. As a first step we compute, at time step $k$, an empirical 90\% quantile of the most recent 20 values $r_{k-19}\clc r_k$. The largest such quantile is of course just the 2nd largest of the 20 values, and this is our predicted quantile $\hat{q}_{k+1}$ for $r_{k+1}$. Perfect calibration would mean that on average the realized value exceeds the predicted quantile 10\% of the time. Figure \ref{cal0} shows the achieved calibration, i.e. graphs
\[ y_k =\frac{1}{k}\sum_{j=1}^k\1_{(r_j> \hat{q}_j)}.\]

\begin{figure}[h]
\centering
\begin{minipage}[b]{0.45\linewidth}
\includegraphics[scale=0.5]{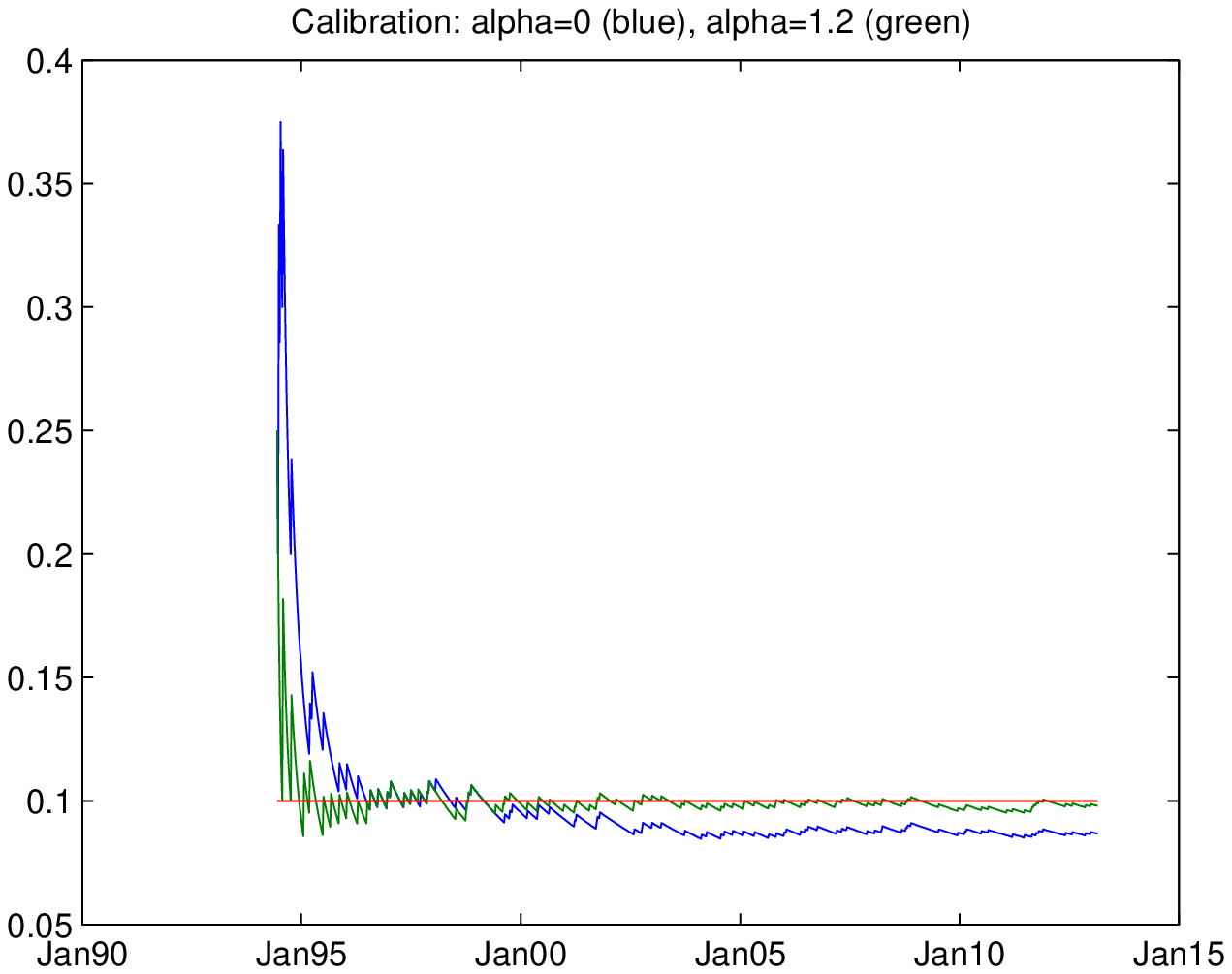}

\vspace{-4mm}
\begin{center}(a) Calibration\end{center}
\end{minipage}
\qquad
\begin{minipage}[b]{0.45\linewidth}
\includegraphics[scale=0.5]{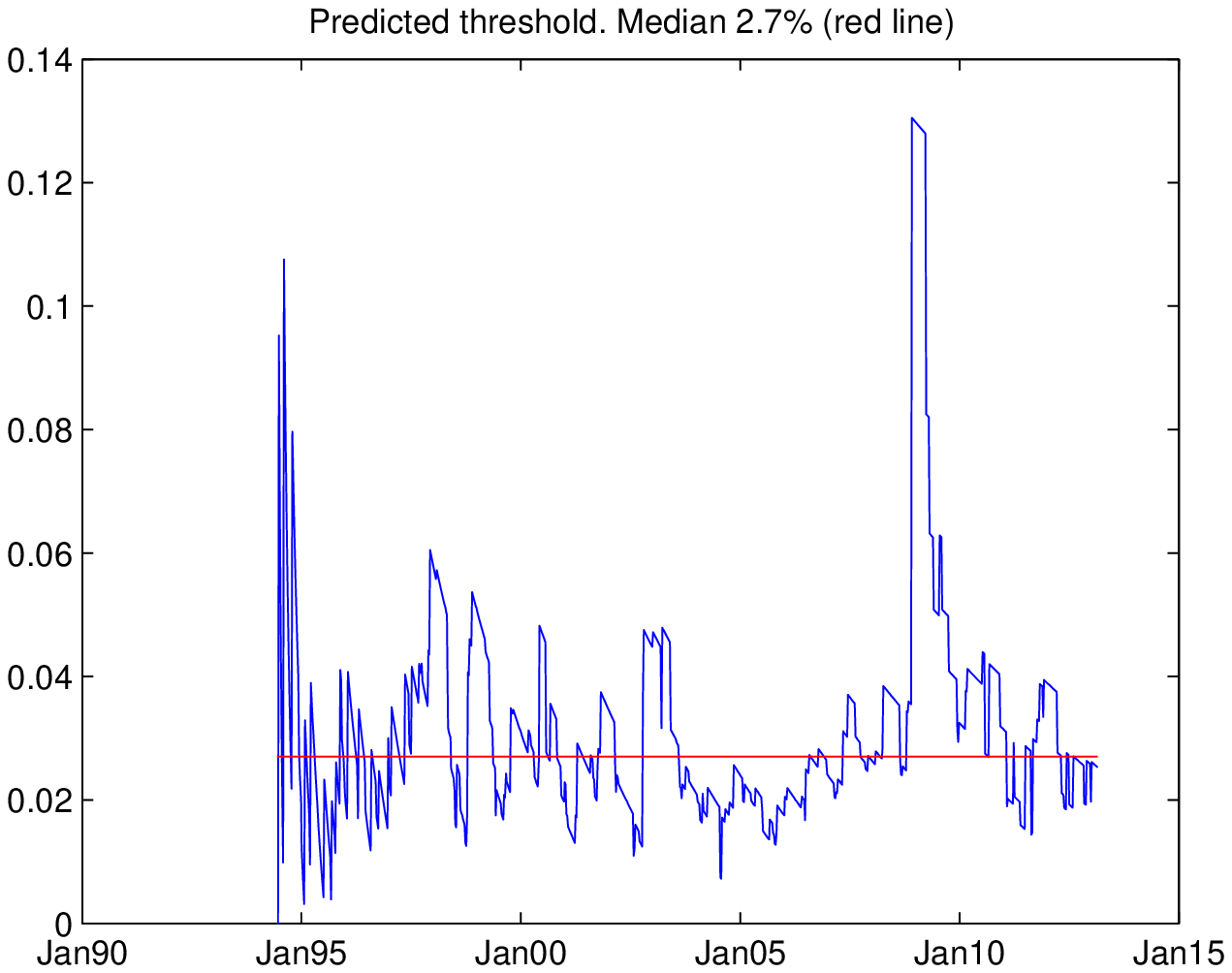}

\vspace{-4mm}
\begin{center}(b) Quantile estimates\end{center}
\end{minipage}
\caption{Performance of feedback algorithm}
\label{fba}
\end{figure}

As can be seen, the algorithm is slightly miscalibrated in that the average exceedance rate is around 9\% rather than the desired 10\%, showing that the average threshold is too high. This may be related to the fact that we chose the largest possible quantile, but in any case is easily rectified by a simple feedback or \emph{adaptive} mechanism that corrects for errors in the achieved performance. Specifically, the new quantile prediction is
\be \check{q}_{k+1}=\hat{q}_{k+1}+\varphi(\check{y}_k-0.1)\label{qp1}\ee
where 
\be \check{y}_k =\frac{1}{k}\sum_{j=1}^k\1_{(r_j> \check{q}_j)}\label{qp2}\ee
and $\varphi$ is a parameter. The performance of this algorithm and the original one are shown in Figure \ref{fba}(a), while Figure \ref{fba}(b) 
shows the sequence of thresholds produced by the algorithm, which vary dramatically over time. (The straight line is the median threshold, around 0.028.) The value of $\varphi$ chosen was 1.2. Performance is not very sensitive to this value, but some number greater than one speeds up convergence of the algorithm.
\begin{table}[tbh]
\begin{tabular}{|c|c|c|c|}\hline
$k$ & Min \%& Max \%&Bernoulli SD\\ \hline
50&8.96&11.22&4.24\%\\
100&9.09&11.22&3.00\%\\
250&9.50&10.53&1.90\%\\
500&9.67&10.33&1.84\% \\ \hline
\end{tabular}
\caption{Calibration performance of adaptive algorithm}
\label{calper}
\end{table}

Table \ref{calper} quantifies the performance of the algorithm. Each row of the table shows, for the number of weeks $k$ in the left-hand column, the minimum and maximum exception frequencies $\min\{\check{y}_k\clc\check{y}_{1500}\}$ and
$\max\{\check{y}_k\clc\check{y}_{1500}\}$, with $\check{y}_j$ given by \eqref{qp2}, experienced beyond that time to the end of the sample. For comparison, the last column shows the standard deviation $\sqrt{p(1-p)/k}$ of the average of $k$ independent Bernoulli trials with success probability $p=0.1$. All numbers are expressed in percentage terms. 
It appears that the deviations from the theoretical 10\% exception frequency are well within the sampling error expected under the i.i.d. hypothesis. Stable calibration is maintained right through the financial crisis period of 2007-09, though of course the actual thresholds fluctuate widely in response to market conditions.

\citet*{kmp06} present a highly informative and extremely thorough comparative study of VaR estimation techniques. They summarize their conclusions as follows:
\begin{quote}In this study we compare the out-of-sample performance of existing methods and some new models for predicting value-at-risk (VaR) in a univariate context. Using more than 30 years of the daily return data on the NASDAQ
Composite Index, we find that most approaches perform inadequately, although
several models are acceptable under current regulatory assessment rules for model adequacy. A hybrid method, combining a heavy-tailed [ .. ] GARCH filter with an extreme value theory-based approach, performs best overall.
\end{quote}
Given this conclusion and the quite high computational demands of the methods surveyed, it does appear from the results presented here that data-driven methods, including techniques such as reinforcement learning \citep{demlee06} merit further investigation.

\subsection{Testing for serial dependence} We now implement the test for serial dependence introduced in Section \ref{ss-sd}. The null hypothesis is
\[\mathfrak{H}_0: \mbox{The $Y_k$ are i.i.d. with $\bP[Y_k=1]=\mu$}.\]
$\mH_0$ is rejected if $\hat{q}_\beta\notin[t_1, t_2]$; these intervals are specified for various significance levels and data lengths in Tables \ref{t1} and \ref{t2}.

\begin{figure}[h!]
\centering
\begin{minipage}[b]{0.45\linewidth}
\includegraphics[scale=0.5]{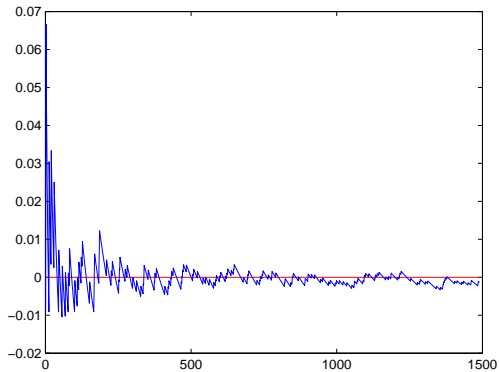}

\vspace{-1mm}
\begin{center}(a) Whole-sample.\end{center}
\end{minipage}
\qquad
\begin{minipage}[b]{0.45\linewidth}
\includegraphics[scale=0.5]{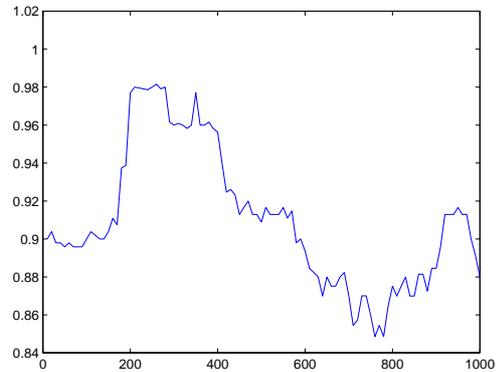}

\vspace{-6mm}
\begin{center}(b) Running calibration $z_k$ of \eqref{qp3}\\ with data window 500.\end{center}
\end{minipage}
\caption{Calibration with $\beta=0.90$.}
\label{90}
\end{figure}

First, we take $\beta=0.9$ as above and run the test on the maximum data length 1500 for the FTSE100 return series. The calibration results, shown in Figure \ref{90}(a) are similar to the shorter data length results of Figure \ref{fba}. For the independence test we determine the relative frequencies $\bar{n_1}, \bar{n_2}$ and then calculate $\hat{q}(\bar{n_1}, \bar{n_2})$. From Theorem \ref{thm-qhat}, the limiting values of $\bar{n_1}, \bar{n_2}$ under $\mH_0$ are $(1-\beta)^2,\, \beta^2=0.01, 0.81$ respectively.
The values obtained in our test were
\bstar \bar{n}_1(1500)&=&0.0100\\
 \bar{n}_2(1500)&=&0.8120\\
 \hat{q}_{0.9}(\bar{n}_1,\bar{n}_2)&=&0.8980.\estar
The agreement with the theoretical limiting values is almost perfect and the value of $\hat{q}_{0.9}$ is within 20 basis points of the correct value under independence. This test is however based on the entire 30-year data sequence and gives us only one estimate. A better evaluation is to take running estimates over shorter periods. Figure \ref{90}(b) shows the results of estimates with a moving window of length 500. This is a plot of $z_k$ against $k$ where
\be z_k =\frac{1}{500}\sum_{j=k-501}^k\1_{(r_j> \check{q}_j)}\label{qp3}\ee
The confidence intervals in Table \ref{t1} show that $\mH_0$ would only occasionally be rejected at the 5\% significance level.

\begin{figure}[h!]
\centering
\begin{minipage}[b]{0.45\linewidth}
\includegraphics[scale=0.5]{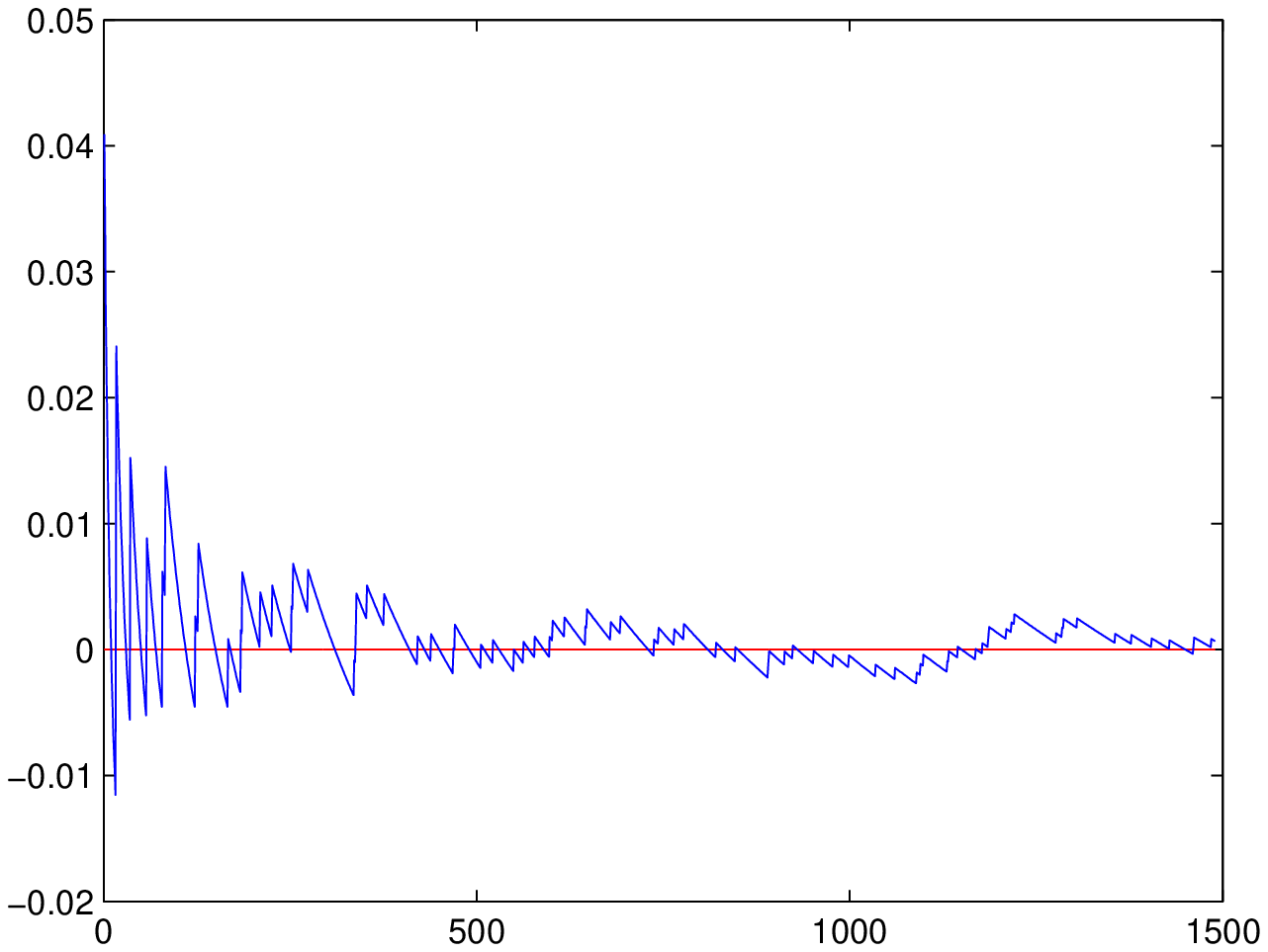}

\vspace{-1mm}
\begin{center}(a) Whole-sample calibration.\end{center}
\end{minipage}
\qquad
\begin{minipage}[b]{0.45\linewidth}
\includegraphics[scale=0.5]{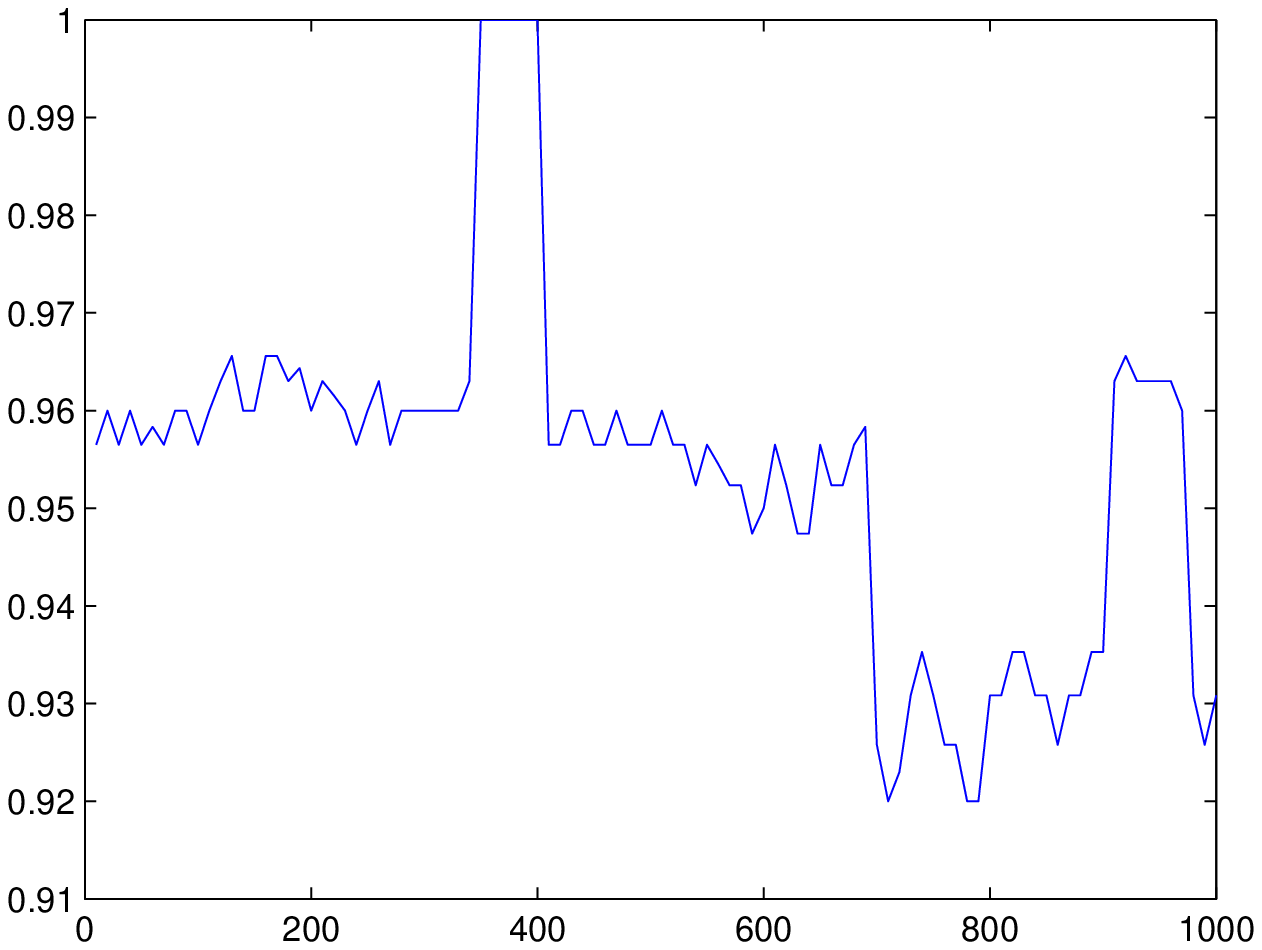}

\vspace{-6mm}
\begin{center}(b) Running calibration $z_k$ of \eqref{qp3}\\ with data window 500.\end{center}
\end{minipage}
\caption{Calibration with $\beta=0.95.$}
\label{95}
\end{figure}

Finally, we repeated these tests at the industry standard value $\beta=0.95$. The results are shown in Figure \ref{95}. The quantile prediction algorithm is the same as before except that we now take the largest, rather than the 2nd largest, of the previous 20 returns as our predictor. Calibration, shown in Figure \ref{95}(a), is only slightly less satisfactory than before. Turning to the independence test, the limiting values of $\bar{n_1}, \bar{n_2}$ in this case are $(1-\beta)^2,\, \beta^2=0.0025, 0.9025$, while the achieved values were
\bstar \bar{n}_1(1500)&=&0.0027\\
 \bar{n}_2(1500)&=&0.9007\\
\hat{q}_{0.95}(\bar{n}_1,\bar{n}_2)&=&0.9481.\estar
The value of $\hat{q}_{0.95}$ is still within 20 bp of the theoretical `independence' value.

 In the running 500 test shown in Figure \ref{95}(b) the estimates never fall outside the range reported in Table \ref{t2}, even at the 50\% significance level. However this test is somewhat less satisfactory in that the upper barrier $t_2$ is always equal to 1 in this case, so the test reduces to a 1-sided one.


\appendix
\section{The Markov chain model}\label{sec:mcm}
In model $\mathfrak{H}_{\mu,\theta}$ the transition probabilities of the chain are as shown in Table \ref{tp}, where $f=(1-\mu)/\mu\leq1$.  The table also indicates the notation $n_i, i=1\clc4$ we use for the number of occurrences of the four pairs $00,11,01,10$ in a sample of size $n$. It should be clear that $n_3$ and $n_4$ play no real role in the problem, since algebraically it must be the case that $|n_3-n_4|\leq 1$, so for a large sample $n_3\approx n_4\approx \frac{1}{2}(n-n_1-n_2)$.

\begin{center}
\begin{table}[hb]
\begin{tabular}[t]{|c|c|l|l|c|}\hline
$x_{k-1}$ & $x_k$ & $p^\theta(x_k|x_{k-1})$&$p^\mu(x_k|x_{k-1})$&\# in sample\\ \hline
0&0&\hspace{2mm}$1-\theta$&\hspace{2mm}$1-\mu$&$n_1$\\
1&1&\hspace{2mm}$1-\theta f$&\hspace{2mm}$\mu$&$n_2$\\
0&1&\hspace{2mm}$\theta$&\hspace{2mm}$\mu$&$n_3$\\
1&0&\hspace{2mm}$\theta f$&\hspace{2mm}$1-\mu$&$n_4$\\
 \hline
\end{tabular}
\vspace{4mm}\caption{\small Markov chain transition probabilities, $f=(1-\mu)/\mu$. The sample size is $n~=~n_1~+~n_2+n_3+n_4$.}
\label{tp}
\end{table}
\end{center}

\vspace{-10mm}\noindent For any $k$ the probability mass distribution of $Y_k$ is $m(x)=1-\mu+(2\mu-1)x,\,x=0,1$, while for $n>0$ the distribution of $(Y_0\clc Y_n)$ is
\[ p^\theta_n(x_0\clc x_n)= m(x_0)\prod_{k=1}^np^\theta(x_k|x_{k-1}).\]
When $\theta=\mu$ the $Y_k$ are i.i.d. with joint distribution $p^\mu_n(x_0\clc x_n)=\prod_0^nm(x_k)$ so, referring to Table \ref{tp}, the likelihood ratio $\mathrm{LR}_n=p^\theta_n/p^\mu_n$ is given by
\bstar \mathrm{LR}^\theta_n(x_0\clc x_n)&=& (1-\theta)^{n_1}(1-\theta f)^{n_2}\theta^{n_3}(\theta f)^{n_4}(1-\mu)^{-(n_1+n_4)}\mu^{-(n_3+n_2)}\\
&=& \{(1-\theta)^{n_1}(1-\theta f)^{n_2}\theta^{(n_3+n_4)}\}\{f^{n_4}(1-\mu)^{-(n_1+n_4)}\mu^{-(n_3+n_2)}\}\\
&=&\{(1-\theta)^{n_1}(1-\theta f)^{n_2}\theta^{n-n_1-n_2}\}\{f^{n_4}(1-\mu)^{-(n_1+n_4)}\mu^{-(n_3+n_2)}\}
\estar
and of course $\mathrm{LR}^\mu_n\equiv 1$. The log likelihood ratio is therefore
\[ \mathrm{LLR}^\theta(n_1\clc n_4)=L^\theta(n_1,n_2)+M(n_1\clc n_4)\]
where
\be L^\theta(n_1,n_2)=n_1\log(1-\theta) +n_2\log(1-\theta f)+(n-n_1-n_2)\log \theta\label{L}\ee
and $M=\mathrm{LLR}^\theta-L^\theta$ does not depend on $\theta$.
\begin{proposition}\label{pA1} For $\beta\geq\ha$, the maximum likelihood estimator is given by
\be \hat{\theta}(\bar{n}_1,\bar{n}_2)= \frac{1}{2f}\left(\bar{n}_{-2}
+f\bar{n}_{-1}-\sqrt{(f-c_1)^2+4f(c_1-c_2)}\right)\label{qhat-app}\ee
where $\bar{n}_{-i}=(n-n_i)/n=1-\bar{n}_i$.
\end{proposition}
\proof To compute the maximum likelihood estimate we maximize $L^\theta$ over $\theta$. We have
\bstar \frac{\d L^\theta}{\d \theta}&=&-\frac{n_1}{(1-\theta)}-\frac{fn_2}{(1-\theta f)}+\frac{n-n_1-n_2}{\theta}\\
&=& \frac{Q(\theta)}{\theta(1-\theta)(1-\theta f)}
\estar
where
\be Q(\theta)=fn\theta^2-(n-n_2 +(n-n_1)f)\theta+(n-n_1-n_2).\label{Qq}\ee
The discriminant of $Q$ is
\bstar D&=&(n-n_2 +(n-n_1)f)^2-4fn(n-n_1-n_2)\\
&=& n^2[(f+c_1)^2-4fc_2],  \estar
where 
\[c_1=1-\frac{n_2+fn_1}{n},\qquad c_2=1-\frac{n_1+n_2}{n}.\]
Under our standing assumption $\mu\geq\frac{1}{2}$ we have $f\leq 1$ and hence $c_1\geq c_2$. Now $D$ can be expressed as 
\[ D=n^2[(f-c_1)^2+4f(c_1-c_2)],\]
showing that $D\geq0$ whatever the values of $n_1,n_2$. Taking into account that $Q(0),Q(1/f)>0$ and $Q(1)<0$ we easily see that $Q$ has a root in each of the intervals $(0,1), (1,1/f)$ so the maximizing $\hat{\theta}\in(0,1)$ is the smaller of the two roots, which is given by \eqref{qhat-app}.
\endproof

\begin{proposition}\label{pA2}
In any model $\mathfrak{H}_{\mu,\theta_0}$ with $\mu\in[\frac{1}{2},1],\, \theta_0\in[0,1]$, the estimator $\hat{\theta}$ is consistent, i.e. $\hat{\theta}(\bar{n}_1,\bar{n}_2)\to \theta_0$ a.s. as $n\to\infty$.
\end{proposition}
\proof Associated to the chain $Y_k$ is the 4-state Markov chain $Y_k, k=1,2,\ldots$ where $Y_k$ takes the values $1,2,3,4$ respectively when
$(Y_{k-1},Y_k)=(0,0), (1,1),(0,1),(1,0)$. The transition matrix for this chain is
\[ \mathbf{P}=\left[\begin{array}{cccc}1-\theta_0&0&\theta_0&0\\0&\theta'_0&0&1-\theta'_0\\0&\theta'_0&0&1-\theta'_0\\1-\theta_0&0&\theta_0&0\end{array}\right] \] 
Consider first the case $\theta_0\in(0,1)$. Then the chain $Y_k$ is irreducible and recurrent and consequently has a unique stationary distribution $\mathbf{m}$ characterized by the property that $\mathbf{m}'(\mathbf{I}-\mathbf{P})=0$, where $\mathbf{I}$ is the $4\times4$ identity matrix. This system of equations is readily solved to give
\[ \mathbf{m} = \left[\begin{array}{c}(1-\theta_0)(1-\mu)\\ \mu-(1-\mu)\theta_0\\ \theta_0(1-\mu)\\ \theta_0(1-\mu)\end{array}\right],\]
where we have substituted for $\theta'_0$ from \eqref{qprime}. The numbers $n_i$ introduced above are simply the numbers of visits to state $i$ by the chain $Y$ in a sample of length $n$. Since $Y$ is recurrent,
\be \lim_{n\to\infty}\frac{n_i}{n} = \mathbf{m}_i\quad\mathrm{a.s.,}\quad i=1\clc 4.\label{lim}\ee
The quadratic form $Q$ of \eqref{Qq} can be written as
\be \frac{1}{n}Q(\theta)=f\theta^2-(1-\bar{n}_2+(1-\bar{n}_1)f)\theta+(1-\bar{n}_1-\bar{n}_2).\label{Qn}\ee
If we substitute $\bar{n}_i=\mathbf{m}_i,\,\,i=1,2$ we find $Q(\theta_0)=0$ and hence that $\hat{\theta}(\mathbf{m}_1,\mathbf{m}_2)=\theta_0$. Now $\hat{\theta}(\bar{n}_1,\bar{n}_2)$ is a continuous function of the two parameters, and hence in view of \eqref{lim} we have $\lim_{n\to\infty}\hat{\theta}(\bar{n}_1,\bar{n}_2)=\theta_0$ a.s.

We now consider the cases $\theta_0=0,1$. When $\theta_0=0$, $Y_k=Y_0$ for all $k$, so either $n_1=n, n_2=0$ or $n_1=0, n_2=n$ giving, from \eqref{L}, values of $L^\theta$ equal to $n\log(1-\theta)$ or $n\log(1-\theta f)$ respectively. In either case, $L^\theta$ is maximized at $\theta=0=\theta_0$.

The case $\theta_0=1$ is a little more tricky. Here $Y_{k-1}=0 \Rightarrow Y_k=1$, so $n_1\equiv 0$. The sample path consists of strings of ones separated by single zeros. The probability of flipping from 1 to 0 is $f$, so the mean length of a string of ones is $1/f$. Each flip from 1 to 0 and back adds 1 to $n_3$ and to $n_4$, and each string of ones of length $m$ adds $m-1$ to $n_2$. So the mean growth rates in $n_3+n_4$ and in $n_2$ are in the ratio $2:(1/f)-1=(2\mu-1)/(1-\mu)$, implying that, loosely speaking, the fraction of the time spent growing $n_2$ is $((2\mu-1)/(1-\mu))/((2\mu-1)/(1-\mu)+2)=2\mu-1$. We conclude that
\[ \lim_{n\to\infty}\bar{n}_2=2\mu-1\quad\mathrm{a.s.}\]
At the limiting value,
\[ \frac{1}{n}L^\theta=(2\mu-1)\log(1-\theta f)+2(1-\mu)\log \theta\]
and the derivative with respect to $\theta$ is 
\[ -\frac{\mu(1-f)f}{1-\theta f}+\frac{2(1-\mu)}{\theta}.\]
This is equal to $+\infty$ at $\theta=0$ and is finite and decreasing for $\theta>0$. Its value at $\theta=1$ is $1-\mu>0$, and we conclude that the maximum occurs at $\theta=1$. A simple continuity argument now shows that $\lim_{n\to\infty}\hat{\theta}(0,\bar{n}_2)= 1=q_0$ a.s.
\endproof
\begin{small}
\bibliographystyle{chicago}

\end{small}

\end{document}